\documentclass[12pt]{article}
\pdfoutput=1
\usepackage{amsmath,amsfonts,amssymb,amsthm,amssymb,bbm,bm}
\usepackage{pdfsync}
\usepackage{mathabx}
\usepackage{verbatim}
\usepackage{graphicx,import}
\usepackage[latin1]{inputenc}
\usepackage{hyperref}
\usepackage{empheq}
\usepackage[all]{xy}
\usepackage{stmaryrd}
\usepackage{rotating}
\usepackage{color}  
\usepackage{slashed}
\usepackage{cancel}
\usepackage{array, makecell} %
\usepackage{comment}
\usepackage{tikz-cd}
\usepackage{hhline}
\usepackage{comment}
\usepackage{cancel}
\usepackage{caption}
\makeatletter

\DeclareFontFamily{U}{matha}{\hyphenchar\font45}
\DeclareFontShape{U}{matha}{m}{n}{
    <5> <6> <7> <8> <9> <10> gen * matha
    <10.95> matha10 <12> <14.4> <17.28> <20.74> <24.88> matha12
     }{}
\DeclareSymbolFont{matha}{U}{matha}{m}{n}
\DeclareMathSymbol{\oright}       {2}{matha}{"69}

\newcommand{\doublehat}[1]{%
\begingroup%
  \let\macc@kerna\z@%
  \let\macc@kernb\z@%
  \let\macc@nucleus\@empty%
  \hat{\raisebox{.55ex}{\vphantom{\ensuremath{#1}}}\smash{\hat{#1}}}%
\endgroup%
}

\newcommand{\p}{\partial}

\newcommand{\bit}{\begin{itemize}}
\newcommand{\eit}{\end{itemize}}
\newcommand{\bd}{\begin{description}}
\newcommand{\ed}{\end{description}}

\newcommand{\bc}{\begin{center}}
\newcommand{\ec}{\end{center}}

\newcommand{\C}{{\mathbb C}}

\newcommand{\R}{{\mathbb R}}
\newcommand{\Z}{{\mathbb Z}}

\newcommand{\cM}{{\cal M}}

\newcommand{\hD}{\widehat\Delta}
\newcommand{\hN}{\widehat{N}}

\newcommand{\cQ}{{\mathcal{Q}}}

\def\be#1\ee{\begin{align}#1\end{align}}
\newcommand{\bea}{\begin{eqnarray}}
\newcommand{\eea}{\end{eqnarray}}
\newcommand{\bs}{\begin{subequations}}
\newcommand{\es}{\end{subequations}}

\newcommand{\la}{\label}

\newcommand{\f}{\frac}

\newcommand{\bz}{{\bar{z}}}
\newcommand{\bw}{{\bar{w}}}

\def\p{\partial}

\def\d{\delta}

\def\rd{\mathrm{d}}
\def\pa{\partial }

\def\k{{\kappa^2} }

\newcommand{\tC}{\widetilde{C}}

\newcommand{\tN}{\widetilde{N}}
\newcommand{\hC}{\widehat{C}}

\newcommand{\scri}{\mathscr{I}}

\usepackage[mathscr]{euscript}

\begin{document}

\begin{titlepage}
\unitlength = 1mm
\ \\
\vskip 3cm
\begin{center}

{\LARGE 
\textsc{A discrete basis for celestial holography}}\par

\vspace{0.8cm}
Laurent Freidel$^1$, Daniele Pranzetti$^{1,2}$,
Ana-Maria Raclariu$^{1,3}$
\vspace{1cm}

{\it $^1$Perimeter Institute for Theoretical Physics, 31 Caroline Street North, Waterloo, Ontario, Canada N2L 2Y5\\ \smallskip
\it $^2$ Universit\`a degli Studi di Udine, via Palladio 8,  I-33100 Udine, Italy\\ \smallskip
\it $^3$ Universiteit van Amsterdam, Science Park 904, 1098 XH Amsterdam, the Netherlands}

\vspace{0.5cm}

\begin{abstract}
Celestial holography provides a reformulation of scattering amplitudes in four dimensional  asymptotically flat spacetimes in terms of conformal correlators of operators on the two dimensional celestial sphere in a basis of boost eigenstates. A  basis of {massless particle} states has been previously identified in terms of conformal primary wavefunctions labeled by a boost weight $\Delta = 1 + i\lambda$ with $\lambda \in \mathbb{R}$. Here we show that a  {\it discrete} orthogonal and complete basis exists for  $\Delta \in \mathbb{Z}$. This new basis consists of a tower of discrete memory and Goldstone observables, which are conjugate to each other and allow to reconstruct gravitational signals belonging to the  Schwartz
space. We show how generalized dressed states involving the whole
tower of Goldstone operators can be constructed and  evaluate the higher spin Goldstone 2-point functions.  Finally, we recast the tower of higher spin charges providing a representation of the $w_{1+\infty}$ loop algebra (in the same helicity sector)  in terms of the new discrete basis.
\end{abstract}

\end{center}

\end{titlepage}

\tableofcontents

\section{Introduction}

Free equations of motion admit highest weight solutions with respect to the Lorentz SL$(2,\mathbb{C})$ that diagonalize boosts towards a point $(z, \bz)$ on the celestial sphere \cite{Pasterski:2016qvg}. These so called conformal primary solutions were shown in \cite{Pasterski:2017kqt} to form a basis for normalizable massless single particle states in four-dimensional (4D) asymptotically flat spacetimes provided the boost weight is in the principal series $\Delta = 1 + i\lambda$ with $\lambda \in \mathbb{R}.$ Any such state can therefore be expressed as a linear combination of operator insertions in 2D celestial conformal field theories (CCFT) living at points $(z,\bz)$ on the celestial sphere and carrying boost weights $\Delta = 1 + i\lambda$.  

While the collection of operators with such dimensions is complete, not all of them need to be in the physical spectrum. In conformal field theory operators typically organize in representations of symmetry algebras \cite{DiFrancesco:1997nk}. 
In CCFT however, basic symmetries such as Poincar\'e translations \cite{Stieberger:2018onx} or conformally soft symmetries \cite{Donnay:2018neh, Adamo:2019ipt, Pate:2019lpp} shift $\Delta$ by positive or negative integers: the basis of operators with $\Delta = 1 + i\lambda$ is not invariant under such transformations. Another example are $d$-dimensional unitary CFTs, where conformal partial waves of dimension $\Delta = \frac{d}{2} + i\lambda$ form a basis of solutions to the two-particle conformal Casimir equation, while physical operators are extracted from expansions in this basis by contour deformation and have real scaling dimensions \cite{Dolan:2011dv,Simmons-Duffin:2012juh}. 
 
Some of the most interesting operators in CCFT are the conformally soft operators \cite{Pate:2019lpp, Fan:2019emx}. In gravity these correspond to conformal primary gravitons of $\Delta = 1 - n$, $n \in \mathbb{N}$. At tree level, these operators generate a $w_{1 + \infty}$ algebra \cite{Guevara:2021abz, Strominger:2021mtt, Himwich:2021dau}. A natural question is whether there exists a basis of conformal primary wavefunctions (or equivalently celestial operators) that includes this tower of conformally soft operators. A positive answer would have remarkable consequences. Perhaps most strikingly, it would imply that at least tree-level higher-point graviton scattering amplitudes are determined recursively from low-point ones by the Ward identities associated with the tower of conformally soft gravitons!

The goal of this paper is to show that conformal primary wavefunctions with $\Delta \in \mathbb{Z}$ form a basis for a special class of ``sufficiently localized'' gravitational signals. In particular we show that news functions that fall off faster than any polynomial at early and late retarded (or advanced) times, and whose Fourier transforms behave similarly,\footnote{Formally, we will restrict our attention to signals in the Schwartz space \cite{Schwartz}.} admit expansions in terms of \textit{either} conformal primary gravitons at negative integer dimensions \textit{or} their canonically conjugate Goldstone partners at positive integer dimensions. We argue that such a restriction on admissible operators can be interpreted as having a both infrared (IR) and ultraviolet (UV) regularized space of states. We show that the reconstruction of the news from these modes then follows from the Ramanujan master theorem \cite{Ramanujan}. As a check of our results, we show that conformal primary wavefunctions at null infinity ($\scri$) are orthogonal and complete with respect to the symplectic form that relates conformally soft (or memory) wavefunctions to their canonically conjugate Goldstone partners.  A similar result was subsequently derived for the two-point function by Cotler, Miller and Strominger in \cite{Cotler:2023qwh}. We derive the symplectic form on the space of discrete conformal primary operators directly from the canonical form on the gravitational phase space \cite{Ashtekar:1978zz}. 

While the semi-infinite tower of conformal primary gravitons at negative integer dimensions has been extensively studied, the properties of their positive dimension counterparts are less understood. The leading Goldstone operators have been first introduced in \cite{Donnay:2018neh}, their relation to infrared dressings was worked out in \cite{Arkani-Hamed:2020gyp}, while aspects of their representation theory were studied in \cite{Donnay:2020guq,Pasterski:2021fjn,Pasterski:2021dqe,Donnay:2022sdg}. In particular, in \cite{Donnay:2020guq}, the positive integer dimension wavefunctions have been argued via an inner product computation to be canonically paired with the negative integer ones. The inner products derived therein depend on a subtle, regularized delta function in dimensions. Instead, here we find simply Kroenecker-delta functions upon taking residues at the negative integer dimensions. Our proof of the completeness of conformal primary wavefunctions at integer dimensions is to the best of our knowledge new. We emphasize that this proof relies crucially on the assumption that the functional space of interest is the Schwartz space.  It is well known that the Schwartz class is dense in the space of $L^2$ normalizable functions and hence its closure coincides with the usual Hilbert space\footnote{We thank Jan de Boer for a discussion on this point.} \cite{Schwartz}. It would be very interesting to understand in detail the physical meaning of this assumption and/or whether it can be lifted.

We discuss two applications of our results, namely the generalization of the Weinberg soft S-matrix \cite{Weinberg:1965nx} to include the whole tower of Goldstone operators and the decomposition of the tower of non-linear gravitational charges derived in \cite{Freidel:2021ytz} in terms of the memory and Goldstone operators. 

It is well known that the soft infrared (IR) divergences in gravity are captured by a correlation function of vertex operators of free bosons (or the leading Goldstone operators) in celestial CFT \cite{Himwich:2020rro, Arkani-Hamed:2020gyp}. Here we propose that part of the IR finite S-matrix can similarly be computed as correlation functions of vertex operators associated with the whole semi-infinite tower of conformal primary gravitons at positive integer dimensions. This calculation relies on the replacement of the soft gravitons involved in ladder exchanges between pairs of external lines by gravitons approximated by the tower of tree-level soft theorems.

The paper is organized as follows. In Section  \ref{sec:Ramanujan} we introduce the discrete memory and Goldstone observables and show how they reconstruct the shear and news signals in the 
Schwartz space, through properties of the Mellin transform and the Ramanujan master theorem. In Section \ref{sec:spot} we characterize the radiative gravitational phase space and the one-particle Hilbert space in terms of these new variables. In Section \ref{sec:basis} we define the associated conformal primary wavefunctions and prove that they form and orthogonal and complete basis. In Section \ref{sec:all-dress} we introduce dressed states involving the whole tower of Goldstone operators, which extend previous constructions involving only the leading terms; we also give a generalized expression for all higher
 Goldstone 2-point functions. In Section \ref{sec:charges} we express the tower of higher spin charges in terms of the new discrete basis; this allows us to write both the soft and hard terms of the charges (including mass and angular momentum),  as a corner $S$ integral, removing the integral over time at $\scri$. 
 Final remarks are presented in Section \ref{sec:conc} and extensive technical material is included in the list of Appendices \ref{App:A}, \ref{app:ramanujan}, \ref{App:symplectic-potential}, \ref{App:R}, \ref{sec:completeness}, \ref{app:dress},  \ref{App:Q}.

\section{Mellin transform and Ramanujan master equation}
\label{sec:Ramanujan}

In this section we review some of the mathematical properties of the Mellin transform and its relation to the Fourier transform. 

Let $C(u)$ be a complex valued function on the real line, with $C^*(u)$ its complex conjugate. 
In gravity, $u$ is to be identified with the retarded time, while $C(u,z)$ promoted to a function of the transverse coordinates  $(z, \bz)$ on the celestial sphere is identified with the gravitational radiation field, or the \emph{shear} projected along unit holomorphic vector fields. We define the \emph{news} field $N(u,z)$ to be the time derivative of its complex conjugate $N(u,z):= \pa_u C^*(u,z)$. Similar statements apply to the electromagnetic radiation field $A(u, z)$ and, more generally, to any spin $s$ asymptotic radiative field.  In the first sections we focus on the evolution of gravitational radiation near $\scri^+$ as a function of retarded time. We take this to be characterized by the signal $C(u)$ and the news $N(u)$ whose dependence on celestial coordinates is irrelevant for this part of the  discussion.

 We denote the Fourier transform of $C(u)$ by
\be
\widetilde{C}(\omega) = \int_{-\infty}^{\infty} du e^{i\omega u} C(u), \quad \omega \in \mathbb{R}.
\ee
Further defining  $\widetilde{C}_+(\omega)$ and $\widetilde{C}_-(\omega)$
\be
\label{Fourier}
\widetilde{C}_+(\omega) &:= \int_{-\infty}^{+\infty} \rd ue^{i\omega u} C(u),
\qquad
\widetilde{C}_-(\omega) := \int_{-\infty}^{+\infty} \rd ue^{i\omega u} C^*(u), \quad \omega > 0,
\ee
where $\widetilde{C}_{\pm}(\omega)$ are positive energy modes at $\mathscr{I}^+$, we have the decomposition\footnote{Upon quantization and for $\omega > 0$, $\widetilde{C}_{+}(\omega)$ and $\widetilde{C}_{-}(\omega)$ respectively create positive and negative helicity gravitons at $\scri^+$.}
\be 
C(u) = \frac{1}{2\pi}\int_0^{\infty} d\omega \left[e^{-i\omega u} \widetilde{C}_+(\omega) + e^{i\omega u} \widetilde{C}_-^*(\omega) \right].
\ee
We  can now introduce the Mellin transforms $\widehat{C}_{\pm}(\Delta)$ of $\widetilde{C}_{\pm}(\omega)$ 
\be 
\label{Mellin}
\widehat{C}_\pm(\Delta) &:= \int_{0}^{+\infty} \rd\omega \omega^{\Delta-1} \widetilde{C}_\pm(\omega).
\ee
Note that \eqref{Fourier} implies that $\widetilde{C}_-(-\omega)= \widetilde{C}_+^*(\omega)$, while $\widehat{C}_\pm(\Delta)$ are mutually independent since the Mellin integrals \eqref{Mellin} are supported on positive energy.
It will also be convenient to introduce the positive energy fields 
 \begin{subequations}\la{C+-}
\be
C_+(u) := \frac1{2\pi} \int_0^\infty \rd\omega   e^{-i\omega u} \tC_+(\omega)
=-\frac1{2i\pi}  \int_{-\infty}^{+\infty} \rd u' \frac{C(u')}{(u'-u+i\epsilon)} \,,\la{C+}\\
 C_-(u) := \frac1{2\pi} \int_0^\infty \rd\omega   e^{-i\omega u} \tC_-(\omega)
=-\frac1{2i\pi}  \int_{-\infty}^{+\infty} \rd u' \frac{C^{*}(u')}{(u'-u+i\epsilon)}\la{C-}\,,\ee 
\end{subequations}
which allow for the decomposition of $C(u)$ as
\be\la{C}
C(u) = C_+ (u) + C_-^*(u). 
\ee  
The different representations $C(u)$, $\widetilde{C}(\omega)$ and $\widehat{C}(\Delta)$ capture the same data but diagonalize different operators: $\pa_\omega = iu$ for the time signal $C(u)$, $\omega =i\pa_u$ for its Fourier transform $\widetilde{C}(\omega)$
and the conformal dimension $\Delta= -\omega \pa_\omega =\pa_u u$ for its Mellin transform $\widehat{C}(\Delta)$.

 We further introduce $N_\pm(u) : =\pa_u C_\pm(u)$, which decomposes, in gravity, the news field $N(u)= N_-(u)+N_+^*(u)$. From these definitions, it immediately follows that
\be 
\label{Nh}
\widetilde{N}_\pm (\omega)=-i \omega \tC_\pm(\omega), \qquad \widehat{N}_\pm(\Delta)= - i   \hC_{\pm}(\Delta+1).
\ee
The Mellin transforms can be directly related to $C(u)$ by\footnote{Here and in the following by $i^\Delta$ and $(-)^\Delta$ we respectively mean $e^{i\frac{\pi}2 \Delta}$ and $e^{i\pi\Delta}$.}
\be 
\hC_+(\Delta) & = \int_{-\infty}^{\infty} \rd u\left( \int_0^{\infty} \rd\omega \omega^{\Delta - 1}  e^{i\omega (u+i\epsilon)} \right) C(u) 
=  i^{\Delta} \Gamma(\Delta) \int_{-\infty}^{+\infty} \rd u  \frac{ C(u)}{(u{+i\epsilon})^{\Delta}}\label{hCu}.
\ee
$\hC_-(\Delta)$ is obtained from the same integral but after replacing $C(u)$ by $C^*(u)$ in the integrand. 
Note that the first equality requires to exchange the $u$ and $\omega$ integrals. This is valid only in the regime where the integrand is such that the integral is absolutely convergent which restricts the regime of validity of \eqref{hCu} to the case where $\Re(\Delta) >0$. 
Other values for $\Delta$ are obtained by analytic continuation. 
For $\Delta =n+1$ with $n$ a positive integer we can explicitly perform the integral by decomposing $C= C_++C_-^*$ into  positive and negative energy contributions as in \eqref{C}. $C_+(u)$ decays in the limit where $\Im(u)\to -\infty$ which implies that we can evaluate the integral by closing the contour in the lower half plane and picking up the contribution from the pole at $u = -i\epsilon$ upon integration by parts.
For $C_-^*$ a similar reasoning requires the contour to be closed in the upper half plane. Since all the poles are in the lower half plane, the contribution from this case vanishes. Overall we find
\be 
\hC_{+}(n+1)=  i^{n+1} n! \int_{L} \rd u  \frac{ C_+(u)}{(u{+i\epsilon})^{n+1}},\label{hCu1}
\ee where $L$ denotes the clockwise contour in the lower half plane.

The Mellin inversion formula takes the form
\bea\label{InverseMel}
\tC_\pm(\omega) = \frac1{2i\pi} \int_{\sigma-i\infty}^{\sigma+i\infty}  \rd \Delta \hC_\pm(\Delta) \omega^{-\Delta}\,,
\eea
for $\omega >0$ and is valid for $\sigma$ in the fundamental strip  provided that $\tC_\pm(\omega)$ is continuous \cite{bertrand:hal-03152634}.
Given $a, b \in \mathbb{R}$, a strip in the complex $\Delta$ plane is defined by $\langle a, b\rangle:= \{ \Delta \in \mathbb{C},a< \mathrm{Re}(\Delta)<b\} $. The fundamental strip of $\hC_\pm(\Delta)$ is the largest strip on which the Mellin transform converges.  
The fundamental strip of $\widehat{C}_\pm$ is $\langle a, b \rangle$ with $b>a$ if 
$\tC_\pm(\omega) = O(\omega^{-a}) $ when $\omega\to 0$ and $\tC_\pm(\omega) = O(\omega^{-b}) $ when $\omega\to \infty$.
In other words, the lower bound of the strip is determined by infrared properties of the radiation field, while the upper bound is determined by its ultraviolet behavior. 
Finally, combining \eqref{InverseMel} together with \eqref{C+-} we can give the time Mellin inversion formula
\be \label{InverseTMel}
C_{\pm}(u) = \frac1{i(2\pi)^2} \int_{\sigma-i\infty}^{\sigma+i\infty}  \rd \Delta \hC_\pm(\Delta) \Gamma(1-\Delta) (iu+\epsilon)^{\Delta-1}.
\ee
To summarise, we have that
the three different representations \cite{Donnay:2022sdg} can be related to each other as follows:\footnote{ Note that the relation between $N(u)$ and $\hN(\Delta)$ is really the difference of two Mellin transforms on $u$, see \eqref{np}.}  
\begin{center}
\begin{tikzcd}[column sep=small]
  \tC(\omega) \arrow[rr, shift left,rightharpoonup,  "\textsf{Energy-Mellin} \eqref{Mellin}"] \arrow[rr, leftharpoondown, "\eqref{InverseMel}" '] \arrow[dr, shift left,rightharpoonup,  "\textsf{Fourier \eqref{Fourier}}" ', near end]
  \arrow[dr, leftharpoondown, "\eqref{C+-}"]
    & & \hC(\Delta) \arrow[dl, shift left,leftharpoonup , "\eqref{InverseTMel}"'] \arrow[dl, rightharpoondown, "\textsf{\eqref{hCu} ``Time-Mellin'' }",near end]\\
 & C(u) &
 \end{tikzcd}\,.
 \end{center}

 In this paper we assume that the fundamental strip of $\widehat{C}$ is $\langle 1, \infty\rangle$ or, equivalently,\footnote{ The notation $\langle 0^+, \infty\rangle$ for $N$ strip means  that the strip is $\langle \epsilon, \infty\rangle$ for an infinitesimal $0<\epsilon<1$.  Under this assumption we can choose $\sigma = 1_+ := \lim_{\epsilon \to 0^+} 1+\epsilon$ in \eqref{InverseMel}. We will show that this choice is necessary in order to have a Hilbert space interpretation. } that the fundamental strip of $\widehat{N}$ is $\langle 0, \infty\rangle$. This amounts to restricting our attention to Fourier signals $\widetilde{C}_\pm(\omega)$ that are exponentially suppressed as $\omega \to +\infty$. The heuristic physical argument for the relevance of such signals relies on the expectation that high energy scattering in gravity is dominated by black-hole production generically leading to a suppression of\footnote{We use the convention $\kappa^2= 32 \pi G$.} ${\cal M}(s,t) \sim e^{-\kappa^2 s/16} $ for the $ 2 \to 2$ scattering amplitude.  As a consequence, it was argued in \cite{Arkani-Hamed:2020gyp} that \emph{ultraviolet (UV) complete} theories of gravity are characterized by the existence of a semi-infinite strip  in the RH complex net boost-weight plane  of 2-2 celestial amplitudes. Our assumption is stronger since it translates into the existence of semi-infinite strips in celestial amplitudes regarded as functions of the individual external dimensions.

We also assume that the theory is \emph{infrared (IR) complete}, characterized by the exponential suppression of  $N(u)$ at early and late retarded times $|u| \rightarrow \infty.$
As shown in \cite{Freidel:2021ytz},
this condition ensures that all higher-spin charges are well defined and leads to the infinite tower of soft theorems encoded in the OPE coefficients of the celestial amplitudes \cite{Strominger:2021lvk,Himwich:2021dau}.
 A heuristic argument for the necessity of late time decay can be linked to the fact that  black holes have to radiate away in the quantum theory according to Hawking \cite{Hawking:1975vcx}. Another heuristic argument for the necessity of an infrared regulator built-in the definition of the Hilbert space goes back to Chung, Kibble, Faddeev and Kulish \cite{Kulish:1970ut} who argued that the dressing of asymptotic states is necessary to relieve the IR divergences.\footnote{ The construction of such an infrared safe Hilbert space was however never achieved. See the recent review \cite{Prabhu:2022mcj}, and references therein.} 
 The necessity of that  assumption relies, as we will see,  on the existence of the higher spin soft charges.

 In mathematical terms, these assumptions amount to considering signals which belong to the Schwartz space $\mathcal{S}$ \cite{Schwartz}.
Physically these imply that the asymptotic states under consideration are \emph{wave packets} which include both a UV and an IR regulator.\footnote{That is wave packets such that $N(u)= O(|u|^{-a})$ and $\widetilde{N}(\omega)= O(\omega^{-a})$ for all $a\in \mathbb{N}$. The Schwartz space can be characterised purely in terms of the time signal as $\pa_u^b N(u)= O(|u|^{-a-b})$ for all $a,b\in \mathbb{N}$. }
 We will see that these assumptions of UV and IR completeness lead to one of the main results of this paper, namely that the phase space of the theory can be characterized by conformally soft modes $\mathscr{M}_\pm(n)$ of negative integral dimensions $\Delta= 1-n$ with $n \in \mathbb{N}$ and their canonically conjugate Goldstone partners $\mathscr{S}_\pm(n)$ of positive conformal dimension $\Delta= 1+n$. Another related result is that we provide a Hilbert space pairing on $\mathcal{S}$ and an isometric embedding of the Schwartz space $\mathcal{S} \to \mathcal{H}$ into the usual one particle Hilbert space. In a language familiar to quantum information \cite{PhysRevA.52.R2493} (see also \cite{Almheiri:2014lwa} for applications to holography)  this means that the UV-IR complete space $\mathcal{S}$ is a code subspace of the usual Hilbert space $\mathcal{H}$.
 The key properties of the Schwartz space are that Fourier transforms preserve the Schwartz conditions on the signal and are also preserved by interactions. That is, the multiplication of two Schwartz signals (potentially involving derivatives) is also in $\mathcal{S}$.

 One might be worried that the hypothesis of infrared completeness is too restrictive. Indeed, Damour \cite{damour1986analytical} was the first to point out in the gravitational context, following an idea similar to  Faddeev--Kulish \cite{Kulish:1970ut, Gomez:2016hxz}, that  massive objects never really decouple in a gravitational theory, no matter how far they are. This means that they never stop radiating at late times through their time-dependent quadrupole. This radiation implies an algebraic decay  $N(u)\sim 1/u^2$, for the shear, which would render it impossible to have any subleading soft theorem!
 This perturbative analysis has been, since then, confirmed in many ways and shown to lead to a failure of the peeling theorem \cite{christodoulou2002global, Kehrberger:2021uvf}. If it survives quantization, this analysis would then undermine the celestial holography project and the construction of higher spin symmetry that we develop here. One way out of this puzzle is to realize that perturbation theory of gravitational systems\footnote{The same analysis would apply in gauge theory.} exhibits tails and memory phenomena \cite{blanchet1987radiative,blanchet2014gravitational,Blanchet:2020ngx}. A deeper analysis of these phenomena reveals that the mere hypothesis that higher order perturbations are smaller than lower order ones fails at late time. Therefore, the late time conclusion reached within perturbation theory is not reliable. This phenomenon is well known in thermal physics under the name of \emph{secular growth}. It usually happens in thermal systems and more generally in any system where the background interaction does not vanish. Recently it was argued in \cite{Burgess:2018sou} that the issue of secular growth is generic in gravitational systems in the presence of black holes.
Overall, this means that the exact status of the decay rate of the news at a late time in a consistent quantum gravity theory is left open and should be seriously studied. 

In this work, we assume that quantum gravity requires the presence of strong asymptotic decay. As a side remark, let's keep in mind that non-Abelian Yang--Mills theory does show the same asymptotic violation of the peeling theorem due to late time interaction at the classical level. But we also expect such a theory to confine at the quantum level, which means that the Yang--Mills radiation field decays exponentially at late time, in agreement with our hypothesis of infrared completeness. 
Finally, it is well-known that the one-particle Hilbert space can be obtained as a closure of the space of Schwartzian states. Therefore if we make the assumption that the S-matrix is a continuous operator then the value of the S-matrix element between arbitrary Hilbert space states should be obtained by taking the appropriate limit of Cauchy sequences of S-matrix elements between Schwartzian states.  In this case  it is not a restriction to look only at the value of the S-matrix on Schwartzian states.  

\subsection{Soft gravitons} \la{sec:Soft}

It is clear from the expression \eqref{hCu} that the Mellin transform $\widehat{N}_\pm(\Delta)$ admits poles on the negative real axis at integer values.
We denote the residues of these poles by  $\mathscr{M}_\pm(n)$
 \be\la{Mn}
 \mathscr{M}_\pm(n) := \mathrm{Res}_{\Delta = -n} \widehat{N}_\pm(\Delta), \quad n \in \mathbb{Z}_+
 \ee
 and we call these observables the higher spin, positive and negative energy \emph{ memory observables}.

 We can obtain the memory observables from  \eqref{hCu} and by using the identity  Res$_{\Delta=-n}\Gamma(\Delta)= \frac{(-1)^n}{n!}$. In order to do so we need to appreciate that the integral involved is obtained  as a limit or by closure into the upper half plane \cite{Belin:2020lsr,Besken:2020snx} (see Appendix \ref{App:A})
 \be \label{softC}
 \mathscr{M}_+(n) 
 : = \lim_{\omega \rightarrow 0^+} \frac{i^n}{n!}\left( \int_{-\infty}^{+\infty} \rd u\,e^{i\omega u} u^n \pa_uC(u) \right) = \frac{i^n}{n!} \left( {\oint}_U\rd u\, u^n \pa_uC_+(u) \right).
 \ee
  Here $0^+$ means we take the limit  $\omega \rightarrow 0$ from above and $U$ is the  upper half plane contour. This limit or contour prescription are needed to ensure the convergence of the integrals and realises the projection onto the positive energy sector.
   $\mathscr{M}_-(n)$ is defined by a similar integral but $C^*(u)$ instead of $C(u)$.
Note that the first memories $\mathscr{M}_\pm(0)$, $\mathscr{M}_\pm(1)$ are also called leading and subleading soft gravitons.\footnote{In gravity we have 
 $\mathscr{M}_{\pm}(n)= { -\frac{\kappa}{4\pi n!}\lim_{\omega \rightarrow 0^+}\p_{\omega}^n\left(\omega a_{\pm}(\omega) \right)}$, where $a_{\pm}$ are graviton modes. {In QED we have $\mathscr{M}_{\pm}^{\rm QED}(n) = -\frac{e}{4 \pi n!} \lim_{\omega \rightarrow 0^+}\p_{\omega}^n\left(\omega a_{\pm}(\omega) \right)$. Note that $\mathscr{M}$ are obtained by contractions with the frame fields like in \cite{Freidel:2021ytz} so any factors $\sqrt{q}$, where $q$ is the determinant of the sphere metric, disappear.}} As we will see, the entire tower of memories determines the hard component of the signal so we find it more suggestive to refer to $\mathscr{M}_{\pm}(n)$ as higher spin memory observables. This derivation means that we can understand 
  $\mathscr{M}_\pm(n)$ as  the Taylor expansion coefficients of $\widetilde{N}_\pm(\omega)$ around $\omega=0$, namely
 \be
 \label{NFourier}
 \widetilde{N}_\pm(\omega)=\sum_{n=0}^\infty  \mathscr{M}_\pm(n) \omega^n, \qquad 
 \mathscr{M}_\pm(n) = \frac{1 }{n!} \left.
 \pa^n_\omega \tN_\pm(\omega)\right|_{\omega=0^+} .
 \ee
In the following it will be convenient  to consider the following  combination of memory observables 
\be 
 \begin{split}
 \mathscr{M}(n) &:=  {{\frac{1}{2}}}\left[i^{-n}\mathscr{M}_+(n) + i^n \mathscr{M}^*_-(n) \right]. \label{Mdef}
 \end{split}
 \ee
 This combination can be simply written as a line integral (no contour involved) 
 \cite{Freidel:2021ytz}\footnote{ This object was denoted $N_s=-\frac{(-1)^s}{2} \mathscr{M}^*(s)$ in \cite{Freidel:2021ytz}. }
 \be \label{SoftC}
 \mathscr{M}(n) =\frac{1}{n!}\left( \int_{-\infty}^{+\infty} \rd u\, u^n \pa_uC(u) \right)
 :=  \lim_{\omega\to 0^+} 
 \frac{1}{2n!}
  \int_{-\infty}^{+\infty} \rd u\,  (e^{i\omega u} + e^{-i\omega u}) u^n  \pa_uC(u) .
 \ee
 In other words $ n!  \mathscr{M}(n)  $ is the $n$-th moment of the map $u\to N(u)$.  These moments exist provided that $N(u)$ decays exponentially at $u\to \pm \infty$. This is our condition of IR completeness.

 Infrared completeness implies that the radiative signal is analytical around $\omega=0$ in Fourier space. As shown in \cite{Freidel:2021ytz} the memories determine the higher spin soft charges through derivatives on the sphere\footnote{\la{foot}These are related to the charges defined in \cite{Freidel:2021ytz} by
 \be
 \label{soft-charges-0}
 q_n^1 
  = \frac{1}{2} (-1)^{n+1} D_z^{n+2}\mathscr{M}^*(n)= 
  -\frac1{4}\left[ i^n q_-^{\rm Soft}(n)  + i^{-n}q_+^{*{\rm Soft}}(n)\right].
 \ee
}
 \be 
 \label{qsoftC}
 q_+^{\rm Soft}(n) = D_{\bz}^{n+2} \mathscr{M}_+(n), \quad  q_-^{\rm Soft}(n) =  D_{z}^{n+2} \mathscr{M}_-(n).
 \ee 
The map $N(u) \to \mathscr{M}(n)$ is a map from radiation data to the soft charge data.
As we will see in more detail later, this map is invertible under the additional condition of UV-completeness. 

It is important to note that the memory observables depend on a choice of cut at $\scri$. Here we have chosen the cut to be at $u=0$. The cut can be shifted to $u=u_0$ by replacing $u^n \to (u-u_0)^n$ in the memory definition \eqref{softC}. This leads to the cut-dependent formula
\be 
\mathscr{M}_\pm(n)[u_0] = \sum_{p=0}^n \frac{(-i u_0)^p}{p!}\mathscr{M}_\pm(n-p).
\ee 
Finally, as shown by  Grant and Nichols \cite{Grant:2021hga} the higher spin memory observables defined here encode the displacement memory effect for geodesics with non-vanishing relative accelerations. More precisely,
given two nearby geodesics we denote the curve deviation observable  by $\Delta \xi(u)=\xi(u)-\xi_{\mathrm{flat}}(u)$, where $\xi =\xi^am_a $ is the geodesic separation at time $u$ projected along the frame\footnote{The frame is   parallel transported along the geodesics.} $m$; $\xi_{\mathrm{flat}}$ is the geodesic separation in flat space  due to the relative initial acceleration. We also denote by $\xi^{(n)}(u)$ the 
 generalized acceleration at time $u$ 
\be 
\xi^{(n)}(u) := \pa_u^n \xi(u).
\ee 
One assumes that the gravitational news $N(u)$ is non-vanishing during an interval $u\in [0,\Delta u]$ and we denote $\xi^{(n)}_{\mathrm{in}}$ the initial generalized acceleration. 
Following \cite{Grant:2021hga} one finds that the geodesic deviation at late time is given in terms of the memory observables by
\be 
\Delta \xi_{\mathrm{out}} =\sum_{n} \frac1{2r} \left[ (n+1)  \mathscr{M}(n) -\Delta u\, \mathscr{M}(n-1)\right] \xi^{(n)}_{\mathrm{in}}.
\ee

\subsection{Goldstone modes}\la{sec:Goldstone}

The  spin $\mp n$ Goldstone operators $\mathscr{G}_{\pm}(n)$ are defined to be the variables conjugated to the soft charges \eqref{qsoftC}.  For our analysis it will be convenient to focus on the corresponding Goldstone gravitons $\mathscr{S}_{\pm}(n)$ which are  defined to be on the other hand the variable conjugated to the memories $\mathscr{M}^*_{\pm}(n)$. Whenever clear from context we will interchangeably refer to $\mathscr{S}_{\pm}$ and $\mathscr{G}_{\pm}$ as Goldstone operators.
The relation between these Goldstone fields is similar to the relation between memories and soft charges, namely
\be\label{Goldstones}
\mathscr{S}_{+}(n) = D_z^{n+2}\mathscr{G}_{+}(n), \qquad 
\mathscr{S}_{-}(n) = D_{\bar{z}}^{n+2}\mathscr{G}_{-}(n).
\ee
 These operators have conformal dimension  $\Delta=1+n$ and can be obtained by evaluating $\widehat{N}_{\pm}(\Delta)$ at positive integer $\Delta$,
\be \label{Softdef}
\mathscr{S}_{\pm}(n) :=\lim_{\Delta \rightarrow n} \widehat{N}_{\pm}(\Delta), \quad n \in \mathbb{Z}_+\,  \quad n \neq 0.
\ee
$\mathscr{G}_{\pm}(0)$ are constituents of the vertex operators which were shown to reproduce the soft S-matrix in gravity in \cite{Himwich:2020rro}, as well as the Faddeev--Kulish dressing responsible for the cancellation of IR divergences \cite{Arkani-Hamed:2020gyp}. We will return to the tower of Goldstone dressings in Section \ref{sec:all-dress}. Note that the definition \eqref{Softdef} is only valid for $n>0$, unless the leading gravitational memory is vanishing. Otherwise, as one can see from \eqref{Mn}, direct evaluation at $\Delta = 1$ will produce a divergent result. An appropriate definition of the leading Goldstone operator was first given in \cite{Donnay:2018neh} by subtracting off this divergence. The degeneracy at $\Delta = 1$ was consequently lifted through the construction of a logarithmic wavefunction canonically conjugate to the $\Delta = 1$ (pure gauge) conformal primary wavefunction. In this paper we will assume that logarithmic modes are absent, although it would be very interesting to include them in the future.  $\mathscr{G}_{\pm}(n)$ for $n \in \mathbb{Z}_+$ sit at corners of the dressing diamonds and are canonically paired with the corresponding entries in the subleading memory diamonds. They were extensively studied, in particular in relation to the leading, subleading and sub-subleading soft graviton theorems, in \cite{Pasterski:2021fjn,Pasterski:2021dqe}. We see from the definition \eqref{hCu}, that the Goldstone operator can be written as the  integral\footnote{We can also express it in terms of Fourier modes as
\be 
\mathscr{S}_{\pm}(n) = -i\int_{0}^{+\infty} \rd\omega \omega^n \widetilde{C}_\pm(\omega).\ee}
\be
\label{S-Cu}
\mathscr{S}_{+}(n) =
i^{n} n!  \int_{-\infty}^{+\infty} \rd u \frac{ C(u)}{(u{+i\epsilon})^{n+1}}.
\ee
$\mathscr{S}_-$ is given by \eqref{S-Cu} with $C(u) \rightarrow C^*(u)$. 
The existence of the Goldstone operators relies on our assumption of UV-completeness that the  fundamental strip for $\widehat{N}$ is $\langle 0, \infty \rangle$.
On the other hand, from the definition \eqref{C+-} of the positive energy field and the relation \eqref{Nh}, we have that 
\be
\label{C+0}
\left.  \pa_u^n C_+(u)\right|_{u=0} = -\frac{n!}{2i\pi}  \int_{-\infty}^{+\infty} \rd u' \frac{C(u')}{(u'+i\epsilon)^{n+1}} =  -\f{i^{- n}}{2\pi i} \mathscr{S}_+(n),
\ee
and similarly for $C_-$ with $C(u)$ replaced by $C^*(u)$. This implies that the Goldstone operators appear as Taylor coefficients in the analytic expansion of $C_\pm(u)$ around $u=0$
\be\label{uexp}
C_\pm (u)
=\f{i}{2\pi } \sum_{n=0}^\infty \frac{(-iu)^n}{n!}   \mathscr{S}_\pm(n)\,.
\ee

To summarize, the UV-completeness condition is equivalent to analyticity of the time signal in $u$ around $u=0$.
It is interesting to note that the Goldstone modes can also be defined at cuts other than $u=0$. The change of cut from $0\to u_0$ is implemented by the transformation
\be
\mathscr{S}_\pm(n) 
\to 
\mathscr{S}_\pm(n)[u_0]=  \sum_{m=0}^{\infty}\frac{(-i u_0)^m}{m!} \mathscr{S}_\pm(m+n).
\ee 
This transformation mirrors the one for the memories. While the time translations of the memories involves lower spin memories, the time translation of the Goldstones involves higher spin \footnote{$\mathscr{G}_{\pm}(n)$ are higher spin while $\mathscr{S}_{\pm}(n)$ are their descendants and have $s = \pm 2$ and higher dimensions.} Goldstones.

\subsection{From soft to hard}

In this section we characterize the class of signals that are completely determined in terms of \textit{either} the $\mathscr{M}_{\pm}(n)$ \textit{or} the $\mathscr{S}_{\pm}(n)$ modes defined in Sections \ref{sec:Soft} and \ref{sec:Goldstone}. As we will see, a sufficient condition for this to be the case is that the signal $N(u)$ belongs to a Schwartz space.\footnote{
The Schwartz space $\mathcal{S}$ is defined \cite{Schwartz} as the set of continuous functions $f(u)$ that decay faster than any positive inverse power of $u$ as $|u| \rightarrow \infty$. Formally,
\be 
\mathcal{S} = \left\{f \in C^{\infty}| \forall \alpha, \beta \in \mathbb{N}, || f ||_{\alpha, \beta} < \infty \right\},
\ee
where $
||f||_{\alpha, \beta} = \underset{u \in \mathbb{R}}{\rm sup} \left|u^{\alpha} \p_u^{\beta} f(u) \right|.$ 
}
An important property of Schwartz spaces is that they are preserved by the multiplication operation  and  the Fourier transform: if a function  $f \in \mathcal{S}$, so is its Fourier transform $\widetilde{f} \in \mathcal{S}$.
Similarly if $f,g \in \mathcal{S}$ so is $fg$.

Any function with compact support in $u$ belongs to $\mathcal{S}.$  Physically, as shown in \cite{Freidel:2021ytz}, the condition that $N(u) \in \mathcal{S}$ implies the existence of higher spin charges governed by the set of recursive differential equations
\be 
\p_u\mathcal{Q}_s = D_z \mathcal{Q}_{s - 1} + \frac{s + 1}{2} C Q_{s - 2}, \quad \mathcal{Q}_{-2} = \frac{1}{2} \p_u N\,.
\ee
The canonical symmetry charges are obtained after renormalization and are given by 
\be
q_s =\sum_{p=0}^s \frac{(-u)^p}{p!} D_z^p \mathcal{Q}_{s-p}.
\ee 
These charges are conserved in the absence of radiation, and they generate  an infinite tower of higher spin symmetries. Moreover their conservation implies the tower of tree-level soft theorems \cite{Guevara:2021abz, Strominger:2021lvk}. 
The existence of these charges requires the infrared finiteness condition on the news.
In particular, $N(u) \in \mathcal{S}$ ensures the existence of the conformally soft operators $\mathscr{M}_{\pm}(n)$ which determine the soft charges as shown in \cite{Freidel:2021ytz} (see \eqref{qsoftC} and  section \ref{sec:renq} for more details).

That signals $N(u) \in \mathcal{S}$ can be reconstructed from  either the memories  or the Goldstone modes follows from analyticity. Indeed analyticity in $u$ means that $N(u)$ is given by the analytic continuation of the resummation of its Taylor expansion. This gives the map $\mathscr{M}_\pm(n)\to N(u)$. Similarly analyticity in $\omega$ gives the reconstruction of $\widetilde{N}(\omega)$ from its Taylor coefficients $\mathscr{M}_\pm(n)$. 
A convenient way to understand the reconstruction of the signal from its Taylor expansion involves the Mellin transform of the signal together with the celebrated  Ramanujan master theorem \cite{RamanujanM}. This theorem states that, given a function $\tilde f(\omega)$, which admits an expansion
\be 
\label{eq:RMT0}
\tilde{f}(\omega) = \sum_{n = 0}^{\infty} F(n) \frac{(-\omega)^n}{n!},
\ee
its Mellin transform is given by\footnote{This theorem provides an efficient way to evaluate Mellin transforms, for instance
\be
\frac{1}{(1+\omega)^\alpha}= \sum_{n=0}^\infty \frac{(-1)^n}{n!} \frac{\Gamma(\alpha + n)} {\Gamma(\alpha)} \omega^n \,\, \to\,\,\int_0^{\infty} \rd\omega \frac{ \omega^{\Delta-1}}{(1+\omega)^\alpha}
= \frac{\Gamma(\Delta) \Gamma(\alpha-\Delta)}{\Gamma(\alpha)},
\ee
which reproduces the classical beta integral. A simple but  formal derivation of this theorem uses that   the LHS of \eqref{RMT} can be written  in terms of $E:= e^{ \pa_\Delta}$ as 
\be
\int_0^{+\infty} \rd\omega \omega^{\Delta-1} \left(\sum_{n = 0}^{\infty} F(n) \frac{(-\omega)^n}{n!} \right)= 
 \int_0^{+\infty} \rd\omega \omega^{\Delta-1} e^{-\omega E} F(0)=
\Gamma(\Delta) E^{-\Delta} F(0) = \Gamma(\Delta)  F(-\Delta ). 
\ee
} \cite{RamanujanM}
\be 
\label{RMT}
\widehat{f}(\Delta) = \int_0^{\infty} \rd\omega \omega^{\Delta - 1}\tilde{f}(\omega) = \Gamma(\Delta) F(-\Delta).
\ee
In other words, the Mellin transform is the analytic continuation of $F(n)$. This analytic continuation is unique provided that $F$ is   analytical in the strip $[0,\infty]$, and that\footnote{This is a sufficient condition \cite{Carlson}. It is the same condition under which Carlson's theorem applies, namely that for $F(x) < C e^{\alpha |x|}$ in the RH complex $x$ plane and if $F(0) = F(1) = \cdots = 0$ then $F(x)$ vanishes identically.} \cite{RamanujanM}
\be 
\label{sufficient}
|F(\Delta)| < c\, e^{\alpha \Re(\Delta)}e^{\beta \Im(\Delta)}, \qquad \beta < \pi,
\ee
where $c>0$, $\alpha \in \mathbb{R}$. Upon closing the contour on the left-hand complex plane, \eqref{sufficient} implies that the contribution from the contour at infinity vanishes and \eqref{eq:RMT0} follows. The condition that $\tilde f(\omega) \in \mathcal{S}$ ensures convergence of the series.

We can now apply the Ramanujan
reconstruction theorem \eqref{RMT} to express $\widehat{N}(\Delta)$ as either the analytic continuation of the memories or the Goldstones.
Starting with the expansion \eqref{NFourier} of $\widetilde{N}(\omega) \in \mathcal{S}$ in terms of memory modes, one finds that
\be 
\widehat{N}_{\pm}(\Delta) = \int_0^{\infty} \rd\omega \omega^{\Delta - 1} \widetilde{N}_{\pm}(\omega) 
= \frac{ \pi}{\sin \pi \Delta} (-1)^{\Delta} \mathscr{M}_{\pm}(-\Delta),
\ee
where we used that $\Gamma(\Delta) \Gamma(1-\Delta)= \frac{\pi}{\sin\pi\Delta}$.

We could alternatively apply \eqref{RMT} to the expansion of $N(u)$ in terms of Goldstone modes implied by \eqref{uexp}. Indeed from (\ref{Nh}, \ref{hCu}) we get that $\hN(\Delta)$ can be written as a difference of Mellin transforms with respect to $u$
\be 
\label{np}
\widehat{N}_+(\Delta)=\Gamma(1+\Delta )\left(  \int_{0}^{+\infty} \rd u u^{-\Delta -1} \left[ i^{\Delta} C_+(u)- i^{-\Delta} C_+(-u)\right] \right) ,
\ee 
where we used the $i\epsilon$ prescription to resolve $i^\Delta = e^{i\frac{\pi}2 \Delta}$. 
Using the expansion \eqref{uexp} and applying the Ramanujan theorem separately for the two terms, we get (see Appendix \ref{app:ramanujan} for details)
\be 
\widehat{N}_{\pm}(-\Delta) = \frac{i}{2\pi}
\Gamma(1+\Delta )\Gamma({-}\Delta) [ i^{2\Delta}-i^{-2\Delta}] 
\mathscr{S}_{\pm}(-\Delta)
= \mathscr{S}_{\pm}(-\Delta).
\ee
This gives us, as expected, that the analytic continuation  of the Goldstone current given by the Ramanujan theorem is simply the Mellin transform $\hN(\Delta)$.

\section{Symplectic form}
\label{sec:spot}

In this section we show that the symplectic form in gravity admits a simple expression in terms of the conformally soft and Goldstone operators $\mathscr{M}_{\pm}(n)$ and $\mathscr{S}_{\pm}(n)$.

We start with the radiative symplectic potential\footnote{For  notational simplicity we define $\int_S:=\int_S \rd^2z \sqrt{q}$, with $q$ the determinant of the 2-sphere metric.
} \cite{Ashtekar:1978zz, Ashtekar:1981sf, Ashtekar:2018lor}  
$
\Theta= \int_S \theta (z,\bar{z})
$  with $\theta$ given by
\be
\label{symplectic-form}
\theta  =   \f2\k\int_{-\infty}^{+\infty} \rd u N(u) \delta C(u)\,,
\ee
where $\kappa=\sqrt{32\pi G}$.
We show in Appendix \ref{App:symplectic-potential} that, after a canonical transformation,  the symplectic potential can be 
written as $\theta=\theta_++\theta_- $, where 
\be
\label{Fourier-symplectic}
\theta_\pm =  \frac1{i\pi \k} \int_0^{+\infty} \rd\omega \omega\left[  \tC_\pm(\omega) \delta \tC_\pm^*(\omega) \right]=
\f2\k\int_{-\infty}^{+\infty} \rd u N_\pm(u) \delta C^{ *}_\pm(u)\,.
\ee
This means that $\tC_\pm(\omega)$ represent the holomorphically  polarized degrees of freedom. After quantization they become  the creation operators for the positive-negative helicity gravitons (or photons) $ \tC_\pm(\omega) =-\frac{i\kappa}{4\pi} a_\pm(\omega)$. 

The same symplectic structure can be expressed in a conformal primary basis. Using the Mellin reconstruction formula \eqref{InverseMel}, we find
\be
\label{Mellin-symplectic}
\theta_\pm 
 &= \frac1{2(i \pi\kappa)^2} \int_{\sigma-i\infty}^{\sigma +i\infty} \rd \Delta  \widehat{C}_{ \pm}(\Delta) \delta \hC^*_{ \pm}(2-\Delta^*)\,.
\ee
Therefore, we see that if one chooses $\sigma =1$, the integration is over $\Delta=1+i \lambda$ with $\lambda \in \mathbb{R}$. This weight is such that 
$2-\Delta^*= \Delta$. In other words, we have that 
\be
\theta_\pm = \frac1{2( i \pi \kappa)^2} \int_{1-i\infty}^{1 +i\infty} \rd \Delta   \hC_{ \pm}(\Delta) \delta \hC^*_{ \pm}(\Delta).
\ee
If we restrict the fields to be in the Schwartzian $\mathcal{S}$ we can also express the symplectic structure in terms of the soft residues and the Goldstones.
Using  \eqref{softC} and  \eqref{uexp} we get that  (see Appendix \ref{App:symplectic-potential})
\be \label{Spot}
\theta_\pm &=  \f2\k\int_{-\infty}^{+\infty} \rd u N_\pm(u) \delta C_\pm^*(u)
=   \f{1}{i\pi\k } \sum_{n=0}^\infty  \mathscr{M}_\pm(n) \d \mathscr{S}^*_\pm(n) \,.
\ee
This result clearly shows that  the conformally soft operators $\mathscr{M}_\pm(n)$ are canonically conjugate to the (complex conjugate) Goldstone operators $\mathscr{S}^*_\pm (n)$. It also shows that the soft operators and Goldstone modes allows, under our assumption of UV and IR completeness, for a complete description of the gravitational phase space.

\subsection{Quantization and commutators}\label{sec:QuCom}
In this section we establish that the memory and Goldstone operators define a {\it discrete} basis of the one-particle Hilbert space. We continue focusing on outgoing relations and will explain the inclusion of incoming modes in sections \ref{sec:basis} and \ref{sec:all-dress}.
From the symplectic structure, we can read off the commutators
of $\mathscr{M}(n,z)$ and $\mathscr{S}(n,z)$. The only non-trivial commutation relations are
\be\la{corn-comm}
[\mathscr{M}_\pm(n,z ),\mathscr{S}_\pm^\dagger(m, z')]= \pi \k   \delta_{n,m}\d^{(2)}(z,z'),
\ee
while all other commutators {involving $\mathscr{M}_{\pm}$ and $\mathscr{S}_{\pm}$} vanish\footnote{{On the other hand, the operators $\mathscr{M}$ and $\mathscr{S}$ are not linearly independent. This implies that the commutators $[\mathscr{S}_{\pm}(m,z), \mathscr{S}^{\dagger}_{\pm}(n,z')]$ and $[\mathscr{M}_{\pm}(m,z), \mathscr{M}^{\dagger}_{\pm}(n,z')]$ may be non-vanishing. We discuss these and evaluate the associated commutators which result in non-vanishing, singular two-point functions of Goldstone and memory modes in sections \ref{sec:lin-dep}, \ref{sec:Gold}.}}
\be\la{corn-comm2}
[\mathscr{M}_\pm(n,z),\mathscr{S}_\mp^\dagger(m,z')]=[\mathscr{M}_\pm(n,z),\mathscr{S}_\pm(m,z')] = [\mathscr{M}_\pm(n,z),\mathscr{S}_\mp(m,z')]=0.
\ee
The vacuum  {at $\mathscr{I}^+$}  is chosen to be such that  
\be
\label{eq:vaccum}
\mathscr{S}_\pm(n,z) |0\rangle =0= \mathscr{M}_\pm(n,z) |0\rangle\,
\ee
and, from the commutation relation \eqref{corn-comm}, it is immediate to see the orthonormality condition
\be
\langle 0|  \mathscr{M}_\pm(n,z) \mathscr{S}^\dagger_\pm(m,z') |0\rangle= \pi \k  \delta_{n,m}\d^{(2)}(z,z')\,.
\ee

This means that we can define the discrete states
\be \la{states}
|n,z, \pm \rangle := \f1{\sqrt{\pi\k}} \mathscr{M}^\dagger_\pm(n,z) |0\rangle, \qquad |\hat{n},z,\pm \rangle := \f1{\sqrt{\pi\k}} \mathscr{S}^\dagger_\pm(n,z) |0\rangle,
\ee  while $\langle n,z,\pm|= \frac{1}{\sqrt{\pi\kappa^2}} \langle 0|   \mathscr{M}_\pm(n,z)$ denotes the conjugate state. These states are orthogonal
\be \la{sprod}
\langle n,z,\epsilon | \hat{m},z',\epsilon'\rangle =\delta_{\epsilon, \epsilon'} 
\delta_{n,m}\d^{(2)}(z,z')= 
\langle \hat{n},z, \epsilon| {m},z',\epsilon'\rangle, 
\ee 
where $\epsilon=\pm$.
We can then introduce the projector  
\be \label{proj}
P = \sum_{n\in \mathbb{N}, \epsilon=\pm}\int_S 
|n,z,\epsilon \rangle \langle \hat{n},z,\epsilon | = \sum_{n\in \mathbb{N},\epsilon = \pm}\int_S 
|\hat{n},z,\epsilon \rangle \langle n,z,\epsilon |\,,
\ee 
which is such that 
\be 
P |n,z, \pm\rangle = |n,z, \pm\rangle, \qquad
P |\hat{n},z, \pm\rangle = |\hat{n},z, \pm\rangle.
\ee 
 Note that the identity in \eqref{proj} between the two ways of representing $P$ can be proven as follows
  \be
  \sum_{n=0}^\infty\left(    \mathscr{S}^\dagger_\pm(n)|0\rangle\langle 0|  \mathscr{M}_\pm(n) \right) 
  &=
 -2i\pi   \int_{-\infty}^{+\infty} \rd u \left( N^{ \dagger}_\pm(u) |0\rangle\langle 0|   C_\pm(u) \right)  
 \cr
 &=
 \sum_{n=0}^\infty\left(    \mathscr{M}^\dagger_\pm(n) |0\rangle\langle 0|  \mathscr{S}_\pm(n) \right)\,,
  \ee
  where the first equality is established upon integration by parts and using the boundary condition  $\left[ C_\pm^\dagger(u)  |0\rangle\langle 0|  N_\pm(u) \right]_{u=-\infty}^{u=+\infty}=0$.

 This equality ensures that $P^\dagger = P$. It also implies that we can use either  the memory states $|n,z\rangle$ or  the Goldstone states $|\hat{n},z\rangle$ as a basis to decompose a Schwartzian state.
 More precisely, if one defines the space $\mathcal{H}_{\mathcal{S}}$ of one-particle Schwartzian states to be given by states of the form 
 \be |\Psi_\pm \rangle =\f1{\sqrt{\pi\k}} \int \rd u \int_S \Psi_\pm(u,z) N_\pm^\dagger(u,z)|0\rangle,
 \ee
 where $\Psi_\pm \in \mathcal{S}$, then $P$ is simply the identity operator on $\mathcal{H}_{\mathcal{S}}$.
 This expression connects with our previous analysis in terms of a classical signal which can be recovered from the action of the shear operator, namely  
 \be 
 \hat{C}_{\pm}(u,z) | \Psi_\pm \rangle = \frac{
 { i}\kappa}{2\sqrt{\pi}}\Psi_{\pm}(u,z) |0\rangle.
 \ee 
 We could also consider the exponentiated coherent state 
 \be
 |e^{ \Psi_\pm} \rangle:= \exp \left( \frac{2}{i \kappa^2}\int \rd u \int_S \Psi_\pm(u,z) N_\pm^\dagger(u,z) \right)|0\rangle,
 \ee
 from which the classical signal can be recovered through the expectation value 
 \be 
 \Psi_\pm(u) &= \frac{\langle e^{\Psi_\pm}| \hat{C}(u) | e^{\Psi_\pm} \rangle }{\langle e^{\Psi_\pm} | e^{\Psi_\pm} \rangle}.
 \ee 
 From now on, we work with the one-particle Schwartzian state.
 We see that the Schwartzian factor $\Psi_\pm$ plays the role of the UV/IR regulated classical signal.
 This means that any Schwartzian state  $|\Psi_\pm\rangle \in \mathcal{H}_{\mathcal{S}}$ can be decomposed in terms of either the memory or the  Goldstone  basis 
 \be 
 \label{Schwartzian}
 |\Psi_\pm\rangle =  \sum_{n \geq 0} \int_S  \Psi_{\pm n}(z) | n,z,\pm\rangle = \sum_{n > 0} \int_S  \Psi_{\pm \hat{n}}(z) |\hat{n},z,\pm\rangle,
 \ee
 where we have 
 \be
 \Psi_{\pm \hat{n}}(z)&= \frac{1}{2\pi} \int_{-\infty}^{+\infty}\frac{(iu)^{{n-1}}}{(n-1)!} \Psi_\pm(u,z) \rd u,\la{Psinhat}
 \\
 \Psi_{\pm n}(z)&= \frac{i^{n + 1} n! }{2\pi} \int_{-\infty}^{+\infty}\frac{ \Psi_\pm(u,z)}{(u+i\epsilon)^{n+1}} \rd u =\f1{2\pi}\int_0^\infty \omega^{n} \widetilde{\Psi}_\pm(\omega,z) \rd \omega.\la{Psin}
 \ee
  It is important to note that the Goldstone state $| \hat{0},z,\pm\rangle$ does not enter the second sum in \eqref{Schwartzian}.  This mode is correlated with infrared divergences and is hence expected to be absent in the expansion of Schwartzian states which are infrared finite. We leave a generalization of the construction herein to non-Schwartzian states to future work, although we anticipate this to be challenging in the light of \cite{Prabhu:2022zcr} where it was shown that no Hilbert space exists in the presence of a leading gravitational memory.

$\mathcal{H}_{\mathcal{S}}$ provides a mathematical representation of the space of UV and IR complete states on which we can define the entire tower of soft charges. As mentioned earlier, it is important to realize that, while the discrete decomposition is valid only for Schwartzian states, we also have that the closure $\bar{\mathcal{S}}$ of the Schwartz space with respect to the Hilbert space metric is the full Hilbert space itself \cite{Schwartz,gelfand1968generalized}. This means in practice that we can approximate any state in the Hilbert space by  Schwartzian states. Therefore, the restriction to this space may not be drastic in the sense that knowing the S-matrix for all possible in and out Schwartzian states allows for a natural extension of the S-matrix to the entire Hilbert space by continuity.


\subsection{Linear dependence of $\mathscr{M}$ and $\mathscr{S}$}
\label{sec:lin-dep}
The previous result suggests that the two discrete bases $\mathscr{S}^\dagger_\pm(n,z) |0\rangle$ and $\mathscr{M}^\dagger_\pm(n,z) |0\rangle$ are not linearly independent. It is important to appreciate that the lack of linear dependency can only be rigorously established for Schwartzian states. Since the discrete states are \emph{not} Schwartzian themselves, the  linear relationship among them can only be formally defined and is singular in the abscence of a regulator. The best we can do is to approximate the discrete basis states as a limit of Schwartzian states.

For instance we can consider the Schwartzian state 
\be\la{Schnhat}
\widehat\Psi_{n,\alpha}(u) := \sqrt{\frac{2\pi}{ \alpha }}\, (i \pa_u)^{n-1} e^{-\frac{u^2}{2\alpha}},
\ee which is well defined for $n > 0$.
The state $|\widehat\Psi_{n,\alpha}\rangle$
 approximates $|\hat{n}\rangle$ when $\alpha \to 0$ and we suppress the $\pm$ helicity labels for clarity. To prove this one shows that 
\be \la{deltahatdef}
\langle m | \widehat\Psi_{n,\alpha}\rangle =\widehat\delta_{mn}(\alpha):= \frac{1}{2\pi} \int_{-\infty}^{+\infty}  \frac{ (i u)^{m-1}}{(m-1)!} \widehat\Psi_{n,\alpha}(u) \rd u
\ee 
approaches $\delta_{nm}$ when $\alpha \to 0$. Therefore $|\hat{\Psi}_{n,\alpha}\rangle = \sum_{m > 0} \delta_{mn}(\alpha)|\hat{m}\rangle \to |\hat{n}\rangle.$ To see this, first note that, after integration by parts, 
$\hat \delta_{nm}(\alpha)=0$ if $n>m$. For general values of $n,m $ we have (see Appendix \ref{App:R}) 
\be\la{deltahat}
\widehat\delta_{nm}(\alpha) 
= \pi_{m-n}\left(-\frac{\alpha}{2}\right)^{\frac{(m-n)}{2}} \f1{\Gamma\left(\frac{m-n}{2}+1\right)},
\ee
where $\pi_{n}=0$ if $n \in \mathbb{Z}$ is negative and if $n$ is odd positive  and $\pi_{n}=1$ if $n$ is even positive or zero. In particular, this means that  \eqref{deltahat} vanishes as $\alpha \rightarrow 0$ unless $m = n$ in which case it is 1.

Similarly, we can construct a state 
$|\Psi_{n,\alpha}\rangle$ which 
 approximates $|{n}\rangle$ when $\alpha \to 0$. We propose the Schwartzian state
\be\la{Psiomega}
 \Psi_{n,\alpha}(\omega) :=2\sqrt{ \frac{2\pi}{  \alpha }}\, \f1{n!}( -\pa_\omega)^{n} \left[c_n\left(1-e^{-\frac{\omega^2}{2\alpha}}\right)^n e^{-\frac{\omega^2}{2\alpha}}\right],
\ee
where $c_n$ is a constant that depends on $n$,
such that the state 
$|\Psi_{n,\alpha}\rangle$
 approximates $|{n}\rangle$ when $\alpha \to 0$.
One can prove this by showing  that 
\be \la{dalpha}
\delta_{nm}(\alpha):=\f1{2\pi} \int_0^\infty \omega^{m} \tilde\Psi_{n,\alpha}(\omega) \rd \omega
\ee 
approaches $\delta_{nm}$ when $\alpha\to 0$. When $m-n<0$ one can show that $\delta_{nm}(\alpha)=0$;
in the case $m-n\geq0$, we can again use integration by part, as the boundary contribution at $\omega=0$ vanishes, 
to compute (see Appendix \ref{App:R})
\be\la{deltaalpha}
\delta_{nm}(\alpha)
&=
\frac{1}{ \sqrt{ \pi  }}
\f{m!c_n}{n!(m-n)!}
 (2\alpha)^{\f{ m - n}{2}}
 \Gamma\left(\f{ m - n+1}{2}\right)
 \sum_{k=0}^n
(-)^k\left (
\begin{matrix}
n \\
k
\end{matrix}
\right)(k+1)^{\f{ n-m -1}{2}}
 \quad \forall~ m-n\geq0,
\ee
which again approaches $\delta_{nm}$ when $\alpha\to 0$, as we fix 
\be
c_n^{-1}=\sum_{k=0}^n
\f{(-)^k}{\sqrt{k+1}}
\left (
\begin{matrix}
n \\
k
\end{matrix}
\right).
\ee

 \subsection{Goldstone and memory 2-point functions}\la{sec:Gold}

{
 The Schwartzian regularization for the discrete basis states \eqref{states} found in the previous section can  be used as a proper regularization for the Goldstone and memory 2-point functions. 
 As we have just seen, two-point functions correspond to the overlap of states such as
\be 
\label{Stwo-pt2}
\langle 0|\mathscr{S}_{\pm}(n, z) \mathscr{S}_{\pm}^{\dagger}(m, w)|0\rangle
&=\pi\k
\langle \hat{n},z, \pm|\hat{m},w,\pm \rangle := \pi \kappa^2 R_{nm} \delta^2(z,w). 
\ee 
The overlap matrix $R_{nm}$
is singular. To see this, one evaluates the correlator 
\be   
\langle 0|C_\pm(u, z) N_{\pm}^{\dagger}(u', w)|0\rangle = \frac{\kappa^2}{4\pi}  \frac{1}{(u-u'-i\epsilon)},
\ee 
where in deriving this result one uses the definitions \eqref{C+-} together with the canonical commutation relations of $\widetilde{C}_{\pm}(\omega), \widetilde{N}_{\pm}^{\dagger}(\omega)$.
Further, using the definition \eqref{C+0} for the Goldstone, one obtains  the unregulated two-point function
\be 
R_{nm}=  \frac{\Gamma(n + m)}{\epsilon^{n + m}},
\ee 
where the limit $\epsilon \to 0$ is implicitely assumed (see Appendix \ref{App:R} for a derivation of this result). 
We would like to interpret this coefficient as evaluating the linear dependence  of the Goldstone state onto the memory one. That is, we expect that 
\be \label{relSM}
\mathscr{S}^\dagger(s,z)|0\rangle  =\sum_{s'} R_{ss'} \mathscr{M}^\dagger(s',z)|0\rangle.
\ee 
As we explained earlier such a relation is formal and should be understood as a limiting relation between Schwartzian states which we now investigate.
}

The regularization of the evaluation \eqref{Stwo-pt2} is done using the Schwartzian states defined before.
It is convenient to introduce the regularized operators
\be  
\mathscr{S}^{\dagger}_{\pm}(\alpha; n, z) := \int_{-\infty}^{+\infty} \rd u \widehat\Psi_{n,\alpha}(u,z) N_\pm^\dagger(u,z), \cr  
\mathscr{M}_{\pm}^{\dagger}(\alpha; n, z) := \int_{0}^{+\infty} \rd \omega  \Psi_{n,\alpha}(\omega ,z) \tilde{N}_\pm^\dagger(\omega,z).
\ee  
 The regularised 2-point functions are then
\be
\label{Stwo-pt}
\langle 0|\mathscr{S}_{\pm}(\alpha;n, z) \mathscr{S}_{\pm}^{\dagger}(\alpha; m, w)|0\rangle
&=\pi\k \langle \widehat\Psi_{n,\alpha}|\widehat\Psi_{m,\alpha}\rangle
 =\pi\k \d^{(2)} (z,w) 
 R_{nm}(\alpha), \\ 
 \langle 0|\mathscr{M}_{\pm}(\alpha; n, z) \mathscr{M}_{\pm}^{ \dagger}(\alpha; m, w)|0\rangle
 &=\pi\k \d^{(2)} (z,w) 
 \widetilde{R}_{nm}(\alpha)\,. \la{Mtwo-pt}
 \ee
 In Appendix \ref{App:R}, using  that the overlap of two Schwartzian states is simply given by
\be 
\langle \Psi_\pm |\Phi_\pm \rangle = \frac{i}{2\pi} \int \rd u \int_S \Psi_\pm^*(u,z) \pa_u \Phi_\pm(u,z),
\ee 
we give the evaluation of the regularized two-point functions 
 \be \la{Rmn}
 R_{nm}(\alpha)= \frac{\Gamma(\frac{m+n}{2})}{\alpha^{\frac{m+n}{2}}}, \qquad \widetilde{R}_{nm}(\alpha) = \frac{C_{nm}}{\alpha^{\frac{m+n}{2}}},
 \ee 
 where $C_{nm}$ is given in \eqref{Cnm}.

 The exact representation of the matrices $R$ and $\widetilde{R}$ depends on the chosen regularization. However, while the limit $\alpha \to 0$ of these matrix elements is singular, we expect the product $\sum_{m} R_{nm}(\alpha) \widetilde{R}_{mp}(\alpha)$ to converge to $\delta_{np}$ in the limit. In this sense, $R$ and $\widetilde{R}$ are approximate inverses of each other.

 Also note that the generic behavior of $R_{nm}$ and $\widetilde{R}_{nm}$ can be derived for a more simple minded approach where we directly regulate the  divergent evaluation 
 \be
 \begin{split}
R_{nm}\delta^{(2)}(z,w)&= \frac{1}{\pi \kappa^2}\langle 0| \mathscr{S}_{\pm}(n,z)\int_0^{\infty} d\omega\omega^{m-1}\sum_{k = 0}^{\infty}\omega^k\mathscr{M}_{\pm}^{\dagger}(k,w)|0\rangle \\
&= \int_{0}^{+\infty} \rd \omega \, 
 \omega^{n+m-1} e^{-\epsilon\omega} \delta^{(2)}(z,w),\,
 \end{split}
 \ee
 where $\epsilon \to 0$ is an ultraviolet regulator\footnote{ The Schwartzian regularisation is of the same nature. As shown in \eqref{Rev}, it corresponds to the insertion of a factor $e^{-\frac{\alpha \omega^2}{2}} $} in this case.
 
Similarly, we would like to show that the second relation in \eqref{Rmn} is consistent with a direct computation of the memory two point function
\be 
\langle 0| \mathscr{M}_{\pm}(m) \mathscr{M}_{\pm}^{\dagger}(n)|0 \rangle.
\ee
This follows from the relation between memory modes and the creation and annihilation operators
\be 
\mathscr{M}_{\pm}(n)=  -\frac{\kappa}{{4}\pi n!}\lim_{\omega \rightarrow 0^+}\p_{\omega}^n\left(\omega a_{\pm}(\omega) \right).
\ee
Specifically, we find 
\be 
\label{m-two-point}
\begin{split}
    \langle 0| \mathscr{M}_{\pm}(m) \mathscr{M}^{\dagger}_{\pm}(n)|0 \rangle &= \frac{\kappa^2}{16 \pi^2 n! m!} \lim_{\omega \rightarrow 0^+} \lim_{\omega' \rightarrow 0^+}\p_{\omega'}^n \p_{\omega}^m \left(\omega \omega' \langle 0|[a_{\pm}(\omega), a^{\dagger}_{\pm}(\omega')]|0 \rangle \right) \\
    &= \frac{i\kappa^2}{2 n! m!} \lim_{\omega \rightarrow 0^+} \lim_{\omega' \rightarrow 0^+} \lim_{\epsilon \rightarrow 0}\p_{\omega'}^n \p_{\omega}^m\left[\frac{\omega'}{\omega - \omega' + i\epsilon} - \frac{\omega'}{\omega - \omega' - i\epsilon} \right] \delta^{(2)}(z, z')\\
    &= \frac{i \kappa^2}{2 n! m!} \lim_{\epsilon \rightarrow 0} n\Gamma(m + n) \left[ \frac{(-1)^{m}}{( i\epsilon)^{m + n}} - \frac{(-1)^{m}}{(- i\epsilon)^{m + n}} \right] \delta^{(2)}(z, z')\\
    & =  \lim_{\epsilon \rightarrow 0} \frac{\kappa^2 i^{m - n +1}}{2 m! (n-1)!}   \frac{\Gamma(m + n)}{\epsilon^{m + n}} \left[1 - (-1)^{m + n} \right] \delta^{(2)}(z, z'),
\end{split}
\ee
where we have used the canonical commutation relations 
\be 
\begin{split}
[a_{\pm}(\omega, z), a^{\dagger}_{\pm}(\omega', z')] &= 2 (2\pi)^2 i \omega^{-1} \lim_{\epsilon \rightarrow 0}\left[\frac{1}{\omega - \omega' + i\epsilon} - \frac{1}{\omega - \omega' - i\epsilon} \right] \delta^{(2)}(z, z')\\
\end{split}
\ee
and adopted the prescription that the $\epsilon \rightarrow 0$ limit is taken at the end. The result has to be symmetric under $n \leftrightarrow m$ so we see that upon symmetrization \eqref{m-two-point} is consistent with \eqref{Rmn}.


\section{A discrete conformal primary basis}\la{sec:basis}

The analysis done in Section \ref{sec:QuCom} can be equivalently carried out at the level of conformal primary wavefunctions. 
Massless scalar conformal primary wavefunctions are given by Mellin transforms of plane waves \cite{Pasterski:2017kqt}
\be
\label{Mellin-cpw}
\varphi^{\pm}_{\Delta}(X; q) =  \int_0^{\infty} \rd\omega \omega^{\Delta - 1} e^{\pm i \omega (q \cdot X \pm i\epsilon)} 
\ee
where $\pm$ distinguish between incoming and outgoing wavefunctions and $q$ is a null vector which we parameterize by points $(z,\bz)$ on the complex plane.
Spinning massless wavefunctions are obtained by dressing \eqref{Mellin-cpw} with frame fields \cite{Pasterski:2020pdk}
\be 
m_{a}^{\pm\mu}(X; q) = \varepsilon^{\mu}_{a} +  q^{\mu}\frac{(\varepsilon_{a}\cdot X)}{(-q\cdot X \mp i\epsilon)}.
\ee
Here $\varepsilon_a^{\mu}$ are polarization tensors defined by
\be 
 \varepsilon_+^{\mu} =  \p_z q^\mu(z, \bz)\,,\qquad \varepsilon_-^{\mu} =  \p_{\bz} q^\mu(z, \bz)
\ee
and obey $\varepsilon_a \cdot \varepsilon_b^* = \delta_{ab}$ and $q\cdot \epsilon_a=0$ provided that
\be \la{qhat}
q(z, \bz) = \frac{1}{\sqrt{2}}\left(1 + z\bz, z + \bz, -i(z - \bz), 1 - z\bz \right).
\ee
It will be convenient to introduce planar retarded coordinates for the Minkowski coordinates
\be 
 X^{\mu}(u,r,w,\bar{w}) = u \p_w\p_{\bar{w}}q^{\mu}(w,\bar{w}) + r q^{\mu}(w, \bar{w}),
  \label{eq:X}
\ee
 in which case case the metric takes the form
\begin{equation} 
\label{eq:planar-Mink}
ds^2 = dX^{\mu} dX_{\mu} = -2 du dr + 2 r^2 d w d\bar{w}.
\end{equation}
With this parameterization $m^{\mu}_a$ take the particularly simple form
\be 
\begin{split}
m^{\pm}_{{a = z};w} &:= \frac{\p X^{\mu}}{\p w} m^{\pm}_{a = z; \mu} = \mp i(\bar{w} - \bz)^2 r^2 \varphi^{\pm}_{\Delta = 1} = (m^{\pm}_{{a = \bz};\bar{w}})^*, \\
m^{\pm}_{a = z;\bar{w}} &:= \frac{\p X^{\mu}}{\p \bar{w}} m^{\pm}_{a = z; \mu} = \pm i ru \varphi_{\Delta = 1}^{\pm} = (m^{\pm}_{a = \bz;w})^*.
\end{split}
\ee
In the large-$r$ limit, $m_{z;w}^{\pm}$ dominate and are given by
\be 
\label{eq:frame}
\lim_{r\rightarrow \infty} m_{z;w}^{\pm} =  -r \frac{\bar{w} - \bz}{w - z}.
\ee
On the other hand at $\scri^+$, adopting the prescription where one first expands the plane wave at large $r$ then evaluates the Mellin transform for generic $\Delta$, \eqref{Mellin-cpw} reduce to \cite{Donnay:2022sdg}\footnote{Shadow contributions do not appear in the order of limits considered here.}
\be 
\label{eq:scpw}
\varphi^{\pm}_{\Delta}(X(u,w); q(z))  \approx  r^{-1}\Gamma(\Delta - 1) (\mp i)^{\Delta-1}( u \mp i\epsilon)^{1 - \Delta} (q^0)^{-\Delta} 2\pi {\delta^{(2)}(z, w)}.
\ee

Relevant here will be the graviton conformal primary wavefunctions\footnote{This differs from the definition in \cite{Pasterski:2020pdk,Donnay:2022sdg} by a factor $\frac{(\pm i)^{\Delta}}{\Gamma(\Delta)}$.}
\be
\label{cpg}
 h_{\Delta, ab;\mu\nu}^{\pm}(X(u, w); q(z)) = m^{\pm}_{(a;\mu} m^{\pm}_{b;\nu)} \varphi^{\pm}_{\Delta}(X; q),
 \ee
 where the brackets denote taking the symmetric traceless combination. We see from \eqref{eq:frame}, \eqref{eq:scpw} that at $\scri^+$ these evaluate to
 \be \la{hpm}
 h^{\pm}_{\Delta,zz;ww} = r \Gamma(\Delta - 1) (\mp i)^{\Delta-1}( u \mp i\epsilon)^{1 - \Delta} (q^0)^{-\Delta} 2\pi {\delta^{(2)}(z, w)} = \left(h^{\pm}_{\Delta^*,\bz\bz;\bar{w}\bar{w}}\right)^*,
 \ee
 with all other components subleading in $r$.
 The operators defined in Section \ref{sec:Ramanujan} can be obtained by computing Klein--Gordon inner products of the quantized metric $h_{\mu\nu}(X)$ with \eqref{cpg}, 
 \be 
 \label{Mellin-wf}
 \begin{split}
 \widehat{C}_{+}(\Delta, z) &= -\lim_{r\rightarrow \infty} \frac{(q^0)^{\Delta} \kappa^2}{4\pi}\langle h_{\Delta; z z}^{-}(X; q)| h(X)\rangle, \\ \widehat{C}_{-}(\Delta, z) &= -\lim_{r\rightarrow \infty} \frac{(q^0)^{\Delta}\kappa^2}{4\pi}\langle h_{\Delta, \bar{z}\bar{z}}^{-}(X; q)|  h(X)\rangle.
 \end{split}
 \ee 
The Klein-Gordon inner product is independent of the slices used to compute it.
At  $\scri^+$ the spin-$2$ generalization of the Klein--Gordon  inner product reduces to
\be \label{KGprod}
\langle h| h' \rangle = -\frac{i}{\kappa^2} \int \rd u \rd^2w r^{-2}\left[h^{AB}\p_u h_{AB}^{'*} - h^{'AB}\p_u h_{AB}^{*} \right],
\ee
where $A, B$ are transverse indices raised/lowered with $ds_{\perp}^2 = 2 \rd z \rd\bz$ and where we recall that $h_{AB} = r C_{AB}.$ Direct evaluation then gives
\be 
\begin{split}
\widehat{C}_+(\Delta, z) &= i^{\Delta} \Gamma(\Delta) \int_{-\infty}^{\infty} \rd u (u + i\epsilon)^{- \Delta} C_{zz}(u,z),\\
\widehat{C}_{-}(\Delta,z) &= (-i)^{\Delta} \Gamma(\Delta) \int_{-\infty}^{\infty} \rd u (u - i\epsilon)^{- \Delta} C_{\bz\bz}(u,z).
\end{split}
\ee
Their counterparts at $\scri^-$ are obtained by instead taking Klein--Gordon inner products with $h_{\Delta;ab}^+.$ 
The rescalings by powers of $q^0$ arise upon conformally mapping the sphere to the plane. To see this recall that the metric on the unit sphere is conformally related to the flat space metric $\rd s^2_{\text{S}}= \frac{2 \rd z \rd \bar{z}}{q_0^2}$.

It is clear from \eqref{hpm} that at $\scri^+$, the net effect of multiplication by the frame fields $m, \bar{m}$ is multiplication by $r$. Therefore, completeness of the scalar conformal primary wavefunctions will imply completeness of the spin-2 conformal primary wavefunctions \eqref{cpg}. 

Our analysis in Section \ref{sec:spot}, and in particular the symplectic potential \eqref{Spot}, suggests that the memory and Goldstone modes
\be 
\label{integer-basis}
\varphi^M_n(X; z) := \underset{\Delta \rightarrow 1 -n}{\rm Res} \varphi^{ +}_{\Delta}(X;\hat{q}), \quad \varphi^G_{n}(X; z) := \lim_{\Delta \rightarrow n + 1} \varphi^+_{\Delta}(X; z), \quad n \geq 0,
\ee
as well as their incoming (-) counterparts, form a basis of scalar conformal primary wavefunctions. We note that the normalization of the memory and Goldstone wavefunctions in \eqref{integer-basis} differs from that in eg. \cite{Donnay:2018neh} by a factor of $\Gamma(\Delta - 1)$. With our normalization, the inner products \eqref{KGprod} produce celestial operators that agree with those obtained by direct Mellin transform of momentum eigenstates. Consequently $\varphi^G$ diverges in the special case $\Delta = 1$. On the other hand, the finite contribution to the same wavefunction,
\be
\varphi^G_0(X;z) := \lim_{\Delta\to 1} [\varphi_\Delta^+(X;z) -\frac1r  \Gamma(\Delta-1) q_0^{-1}2\pi \delta^{(2)}(z,w)] 
\ee
is logarithmic and we expect it to agree with the large-$r$ expansion of the logarithmic modes constructed in \cite{Donnay:2018neh}.\footnote{Note that in contrast to \cite{Donnay:2018neh}, our notation is such that inner products with $M$ and $G$ wavefunctions insert conformally soft and Goldstone operators respectively.} We leave a complete analysis of the discrete basis including the $\Delta = 1$ logarithmic wavefunctions to future work.

In the next section we show 
 that the memory and Goldstone wavefunctions obey orthogonality and completeness relations. This is expressed as the property that  the Klein--Gordon inner product for the wavefunctions \eqref{integer-basis} is given by
\be \label{orthogonality}
\langle \varphi_n^M(X;z)| \varphi_{m}^G(X;z') \rangle = 2(2\pi)^3 \delta_{mn} (q^0)^{-2} \delta^{(2)}(z, z'). 
\ee
It also means that we have the following completeness relation  
\be 
\label{completeness}
\sum_{n \in \mathbb{Z}_+} \int_S \left[ \varphi_n^M(X; z) \varphi^{G*}_{n + 1}(X';z) + (X \leftrightarrow X') \right]= -\frac{(2\pi)^3}{ (q^0(w,\bar{w}))^{3}} \delta^{(3)}(X - X').
\ee
The shift in labels necessary to establish completeness is the discrete analog of the time  derivative entering the Klein--Gordon product.

\subsection{Orthogonality and Completness}

The orthogonality follows from the representation of the conformal primaries \eqref{Mellin-cpw} as Mellin transforms of plane waves together with 
\be 
\langle e^{\pm i \omega \hat{q} \cdot X}| e^{\pm i \omega' \hat{q}' \cdot X}\rangle = \pm 2 (2\pi)^3\omega^{-1} (q^0)^{-2} \delta(\omega - \omega') \delta^{(2)}(z, z').
\ee
Using the defining relations \eqref{integer-basis}, we then have 
\be 
\label{cpw-ip}
\langle \varphi_n^M(X;z)| \varphi_{m}^G(X;z') \rangle &= \lim_{\epsilon \rightarrow 0}\lim_{\Delta \rightarrow 1 - n}(\Delta + n - 1) \int_0^{\infty} \rd\omega  \omega^{- n - 1 + m} 2 (2\pi)^3 (q^0)^{-2} \delta^{(2)}(z, z') e^{-\omega \epsilon}\cr
&= 2 (2\pi)^3\lim_{\epsilon \rightarrow 0}  \epsilon^{n - m} \lim_{\Delta \rightarrow 1 - n} (\Delta + n - 1) \Gamma(m - n) (q^0)^{-2}\delta^{(2)}(z, z') \cr
&= 2 (2\pi)^3 \delta_{mn} (q^0)^{-2} \delta^{(2)}(z, z').
\ee
Since the memory and Goldstone modes are proportional the plane wave modes, with proportionality factor $\widetilde{C}_{+}(\omega) = \frac{\kappa}{4i\pi}a_+(\omega)$, we see that \eqref{cpw-ip} is consistent with \eqref{corn-comm}. 

In the last line of \eqref{cpw-ip} we used that  the discrete delta function admits the following representation in terms of the Mellin transform
\be 
\begin{split}
\label{discrete-df}
\lim_{\epsilon \rightarrow 0}\lim_{\Delta \rightarrow 1 - n}(\Delta + n - 1) \int_0^{\infty} \rd\omega \omega^{\Delta + m - 2} e^{-\omega \epsilon} &= \lim_{\epsilon \rightarrow 0} \lim_{\Delta \rightarrow 1 - n}(\Delta + n - 1) \epsilon^{-\Delta - m + 1} \Gamma(\Delta + m - 1)\\
= \lim_{\epsilon \rightarrow 0} \epsilon^{n - m} \underset{\Delta = 1 - n}{\rm Res}\Gamma(\Delta + m - 1) &= \begin{cases} 0, \quad m \neq n \\
1, \quad m = n.
\end{cases}
\end{split}
\ee
For the last equality on uses that the residue vanishes if $m>n$ while the limit vanishes if $n>m$.

 To prove the completeness relation \eqref{completeness}, we start with the massless  conformal primary wavefunctions evaluated on $\scri^+$
\be
\varphi^+_{\Delta}(X(u,w); \hat{q}(z)) = \frac{\Gamma(\Delta - 1)}{(i u +\epsilon)^{\Delta-1}} \frac{2\pi}{r} (q^0)^{-\Delta} \delta^{(2)}(z, w).
\ee 
Then we show that 
when acting on functions belonging to the Schwartz space, we find that the series
\be 
\label{delta-rep1}
 \frac{1}{2i\pi} \sum_{n \in \mathbb{N}} \left(\frac{u - i\epsilon}{u' + i\epsilon} \right)^n (u' + i \epsilon)^{-1} + (u \leftrightarrow u') =  \delta(u - u'),
\ee
is equal to a  delta distribution.\footnote{The identity $\sum_{n\in \mathbb{Z}} \left(\frac{z}{w}\right)^n =\delta\left(\frac{z}{w}\right)$ is also encountered as a formal power series identity identity in the study of vertex operator algebra \cite{frenkel1989vertex}.  Here we prove it as a distributional identity  on the real line.}
This is proven in Appendix \ref{sec:completeness}.

We then  evaluate 
\be 
\begin{split}\label{Proof}
&\sum_{n \in \mathbb{Z}_+} \int_S\left[ \varphi_n^M(X; z) \varphi^{G*}_{n + 1}(X';z) + (X \leftrightarrow X')\right]\\
&= (2\pi)^2 r^{-2} (q^0)^{-3}\delta^{(2)}(w, w') \left( i\sum_{n \in \mathbb{Z}} \left(\frac{u-i\epsilon}{u' + i\epsilon}\right)^{n}\frac{1}{u' + i\epsilon} + (u\leftrightarrow u') \right) \\
&= -(2\pi)^3 r^{-2} (q^0)^{-3}\delta^{(2)}(w, w') \delta(u - u'),
\end{split}
\ee
where we have used the symmetry under $u \leftrightarrow u'$ to extend the summation range as well as \eqref{delta-rep1}. 
 We can finally check the completeness relation by taking the KG inner product of both sides of \eqref{completeness} with $\varphi_{m + 1}^{M*}$. We find
\be 
\begin{split}
2(2\pi)^3 (q^0)^{-2} \varphi_m^M(X; z) &= -2 (2\pi)^3 i\int \rd u' \rd^2 w'  \delta(u' - u) (q^0)^{-3}\delta^{(2)}(w,w') \p_{u'}\varphi^{M}_{m + 1}(u', w'; z) \\
&= 2(2\pi)^3 (q^0)^{-2} \varphi_m^{M}(X; z),
 \end{split}
\ee
as expected. We used that $ 
\p_u \varphi^M_{m + 1} = i q^{0} \varphi^M_m.$

\section{All order dressing}
\label{sec:all-dress}
In this section, we construct dressed states that diagonalize the tower of soft charges \eqref{softC}. 
We continue working with the planar parameterization \eqref{qhat} and \eqref{eq:planar-Mink}. This means in particular, that we can replace all covariant derivatives $D_z$ by partial derivatives $\pa_z$ in our expressions for the charges and their action.
\subsection{All spin dressing operator}
To construct these states we introduce the Goldstone operators 
\be \label{Goldstones2}
\mathscr{S}(s, z) = \pa_z^{s + 2} \mathscr{G}(s,z),\qquad 
\mathscr{G}(s,z) =\int \rd^2w \, \mathscr{S}(s, w) G_{s+2}^+(w,z).
\ee
 Where is the soft operator \eqref{Softdef} and  $G_{s+2}^+$ is the spin $s$ propagator defined by 
\be  
G^+_{s+2}(w;z) := \frac1{2\pi (s+1)!}\frac{(z-w)^{s+1}}{\bz -\bar{w}},
\qquad \pa_{z}^{s+2} G^{+}_{s+2}(w;z)= \delta^{(2)}(z, w).
\ee
In the following we denote $G^-_{s+2}(w;z)= [G^+_{s+2}(w;z)]^*$ the inverse of $\pa_{\bz}^{s+2}$.

The Goldstone operators can be decomposed into positive and negative helicity states according to 
\be 
\mathscr{G}(s) := {i^{-s}}\mathscr{G}_+(s) - {i^{s}} \mathscr{G}_-^\dagger(s).
\ee
This decomposition is similar to the one given in  \eqref{Mdef} for $\mathscr{M}(n,z)$,
and it implies, given \eqref{corn-comm}, the commutation relations 
\be 
[\mathscr{M}(s,z), \mathscr{G}^\dagger(s',z')] = \pi \kappa^2 \delta_{ss'} G^-_{s+2}(z,z'), \qquad [\mathscr{M}(s,z), \mathscr{G}(s',z')] = 0.
\ee
The Goldstone operators \eqref{Goldstones2} are necessary to define dressed states  as  coherent states 
\be 
\label{eq:Dress}
\begin{split}
    \langle p| \mathcal{D}(p) :=  \langle p|\exp  \left\{\sum_{s = 0}^{\infty} \frac{(-1)^s}{\pi \kappa^2} \int d^2z \left[(-1)^{s} m_-^{(s)}(z;p)\mathscr{G}(s,z) -  m_+^{(s)}(z;p)\mathscr{G}^\dagger(s,z) \right] \right\},
\end{split}
\ee
where $\langle p|$ is an outgoing asymptotic momentum eigenstate and $\mathcal{D}(p)$ is the corresponding  dressing operator. 
 Here  $m^{(s)}_\pm(z;p)$ is a differential operator acting on the momentum labels  that represents the eigenvalue of the hard charge\footnote{
$ q_s^{2}$ was defined in \cite{Freidel:2021dfs, Freidel:2021ytz} and is involved in the derivation of the tower of soft theorems from conservation laws associated with the truncated, renormalized charges $q_s = q_s^{1} + q_s^{2}$. See also footnote \ref{foot} for the normalization of the soft charge.} operator $q^{2}_s(z) $ on asymptotic momentum eigenstates states of positive and negative helicity,  namely\footnote{See equation (84) in \cite{Freidel:2021ytz}}
\be
[ a_{\pm }(p),q_s^{2}(z)] & = \frac{i^{s}}{2} m_\pm^{(s)}(z;p) [a_{\pm }(p)],
\ee
where $m_\pm^{(s)}(z;p)$ is the distributional differential operator  given,
when $p=\omega {q}(w)$, by  \cite{Guevara:2019ypd, Freidel:2021ytz}
\be 
\label{eq:msout}
m^{(s)}_\pm(z;\omega {q}(w)):= \frac{\kappa^2}{4} 
 \omega^{1-s } \sum_{\ell = 0}^s (-1)^{ \ell -s}
\frac{( \ell +1)(\Delta \pm 2)_{s - \ell}}{(s - \ell )!}   \p_{z}^{s - \ell} \delta^{(2)}(z, w) \p_{w}^\ell,
\ee 
where $\Delta= -\omega \pa_\omega$ is the conformal dimension operator and $(x)_n $ is the descending factorial.
For instance, for  the leading, subleading and sub-subleading dressing we  have 
\be 
m^{(0)}_\pm(z;\omega q(w))& =
\frac{\kappa^2}{4} 
\omega \delta^{(2)}(z,w),\cr 
m^{(1)}_\pm(z;\omega q(w))& =
\frac{\kappa^2}{4} \left(
 2 \delta^{(2)}(z, w) \p_{w} -
(\Delta \pm 2)   \p_{z} \delta^{(2)}(z, w) \right), \cr 
m^{(2)}_\pm(z;\omega q(w))& =
\frac{\kappa^2}{4} \frac{1}{\omega} \left(
 3 \delta^{(2)}(z, w) \p_{w}^2 -
2 (\Delta \pm 2)   \p_{z} \delta^{(2)}(z, w) \p_w
+ \frac{(\Delta \pm 2)(\Delta\pm 2 -1)}{2} \p_{z}^2 \delta^{(2)}(z, w) \right).\nonumber
\ee 
We see that only the first two dressing operators are infrared dressings, while  for spins higher than 2 they contain negative powers of $\omega$.
We also see that the spins  correspond to the order of the differential operator.

We demonstrate in appendix \ref{sec:dress-unitary} that the dressing \eqref{eq:Dress} is a unitary operator owing to the non-trivial hermiticity properties of the differential operators $m_{\pm}^s$ defined in \eqref{eq:msout}. In particular its action in a position basis is given by 
\be 
\label{eq:msDout}
i^{1-s} m^{(s)}_\pm(z;w)[f]:= \frac{\kappa^2}{4} \sum_{\ell = 0}^s (-1)^{ \ell -s}
\frac{( \ell +1)(\widehat\Delta \pm 2)_{s - \ell}}{(s - \ell )!}   \p_{z}^{s - \ell} \delta^{(2)}(z, w) \p_{w}^\ell \pa_u^{1-s}f(u,w,\bar{w})
\ee 
with 
$
\widehat\Delta := u\p_u + 1$  
 which becomes diagonal in a conformal primary basis.
We demonstrate in the appendix that 
\be 
\label{eq:hermiticity}
\begin{split}
\langle f | m_+^{(s)}(z;w)[g] \rangle  
&= (-1)^s \langle  m_-^{(s)}(z;w)[f] |g \rangle.
\end{split}
\ee
where $\langle f|g\rangle$ denotes the Klein-Gordon inner product \eqref{KGprod}.

\subsection{Soft theorem and dressing}
In this section we  show that the dressing defined  previously  is the  appropriate one which trivializes the  soft theorems. 
As is well known \cite{Strominger:2017zoo}, the soft theorems are expressed as an operatorial identity on the S-matrix $\mathcal{S}$ 
\be \label{SoftTfull}
\langle {\rm out}| [q_s(z), \mathcal{S}]  |{\rm in} \rangle =0.
\ee
Expanding the charges in terms of their degree in the shear we obtain $ q_s = q_s^1+q_s^2+\cdots$. If we truncate the soft theorems to the quadratic expansion of charges we obtain the tree level, higher spin,  soft theorems. They can be naturally written in terms of the operators\footnote{The normalization is such that $S^{(s)}_-(z,p)$ corresponds to the soft insertion $-\frac1{s!} \pa_\omega^s[\omega a(\omega q(z))]_{0+}$. For $s=0$ this corresponds to Weinberg's soft limit, when $q=\omega q(z)$: \be
\langle {\rm out}| a_{\rm out}(q) S |{\rm in} \rangle \sim \frac{\kappa}{2}\sum_{i} \frac{(p_i\cdot \epsilon)^2}{p_i\cdot q} \langle {\rm out}| S |{\rm in} \rangle
= -\frac{\kappa}{2} \,\sum_{i}\frac{\eta_i \omega_i}{\omega}  \frac{(z_i-w)^2}{|z_i-w|^2}\langle {\rm out}| S |{\rm in} \rangle.
\ee }
\be \label{Sdef}
\frac{\kappa}{4\pi} S^{(s)}_-(z',p) &:= \int \rd^2z   G^+_{s+2}(z,z') m_-^{(s)}(z;p),\cr 
 \frac{\kappa}{4\pi} S^{(s)}_+(z',p) &:= \int \rd^2z   G^-_{s+2}(z,z') m_+^{(s)}(z;p).
\ee  
It was shown in \cite{Freidel:2021ytz} that \eqref{SoftTfull} implies\footnote{Following \cite{Freidel:2021ytz} to fix the normalization, we  have that 
\be
\langle {\rm out}|q_s^1(z) \mathcal{S} |{\rm in}\rangle &= -\frac{i^s}{4} \pa_z^{s+2} \langle {\rm out}| \mathscr{M}_-(s,z)\mathcal{S} |{\rm in}\rangle\cr
\langle {\rm out}|[q_s^2(z), \mathcal{S}] |{\rm in}\rangle &= \frac{i^s}{2} \sum_i m^{(s)}(z,p_i) \langle {\rm out}| \mathcal{S} |{\rm in}\rangle.
\ee
Composing the identity \eqref{SoftTfull} with the propagator $G_{s+2}(z,w)$ therefore gives us \eqref{SoftT}
}the tower of tree level soft theorems \cite{Guevara:2021abz}\footnote{This follows from equation (87) in \cite{Freidel:2021ytz}.}
\be \label{SoftT}
\langle{\rm out}|[\mathscr{M}_-(s,z), \mathcal{S}]|{\rm in}\rangle &\approx \frac{\kappa}{4\pi} \sum_{i = 1}^n S^{(s)}_{-,i}(z)  \langle {\rm out}| \mathcal{S}  |{\rm in} \rangle \\
&= \frac{\kappa^2}{8\pi } 
\sum_{i = 1}^n(\eta_i\omega_i)^{-s + 1} \sum_{\ell = 0}^s \frac{(-1)^{\ell-s}}{\ell!} \frac{(2h_{i})_{s - \ell}}{(s-\ell)!}   \frac{(z - z_i)^{\ell + 1}}{\bz - \bz_i} \p^\ell_{z_i} \left[ \langle {\rm out}| \mathcal{S} |{\rm in} \rangle\right],\nonumber
\ee
 where $\approx$ means that the equality is valid at tree level up to non-universal corrections, and we use the shorthand notation $S^{(s)}_{\pm i}(z)=S^{(s)}_{\pm }(z,p_i) $.
We also denote $\eta_i=+1$ for out states and $\eta_i=-1$ for in states. 
Here $\mathscr{M}_-$ creates a negative helicity graviton in the out state and $n$ is the number of asymptotic hard particles. For the other helicity we exchange $h\to \bar{h}$, $z\to \bar{z}$ which leads to $S_+^{(s)}\to S_-^{(s)}$ where
\be \la{hi}
2h_{i} = -\omega_i \p_{\omega_i} + s_i, \quad
{2\bar h_{i} = -\omega_i \p_{\omega_i} - s_i}.
\ee

The dressings \eqref{eq:Dress} induce shifts in the eigenvalues of the soft gravitons $ \mathscr{M}(s,z)$, $ \mathscr{M}^\dagger(s,z)$,
namely
\be 
[\mathcal{D}(p), \mathscr{M}^\dagger(s',z')] &= 
(-1)^{s}\frac{\kappa}{4\pi}  \mathcal{D}(p)\, S^{({s'})}_-(z',p) ,\\
[\mathcal{D}(p), \mathscr{M}(s', z')] 
&= \frac{\kappa}{4\pi}\mathcal{D}(p)\, S^{({s'})}_+(z',p),\la{dressingid}
\ee 
which imply that, when used as asymptotic scattering states, the dressed states \eqref{eq:Dress} are eigenstates of $q^{\rm Soft}, \bar{q}^{\rm Soft}$ \eqref{softC}. 
This property means that the operators $\mathcal{D}(p)$ are generalizations of the unitary Faddeev--Kulish operators  \cite{Chung:1965zza,Kibble:1968oug,Kibble:1968npb,Kibble:1968lka,Kulish:1970ut}. They are determined by the requirement that they diagonalize the tower of soft operators $\mathscr{M}(s,z)$ together with unitarity.

The dressed amplitude is defined by 
\be {}_{\rm dr}\langle {\rm out}| \mathcal{S} | {\rm in}\rangle_{\rm dr} :=  \langle {\rm out}|  \prod_{i \in {\rm out}} \mathcal{D}(p_i) \mathcal{S} \prod_{j \in {\rm in}}\mathcal{D}^{\dagger}(p_j)| {\rm in}\rangle.
\ee
It is important to note that the tower of soft theorem 
 identities
\be 
{}_{\rm dr}\langle{\rm out}|[\mathscr{M}_-(s,z), \mathcal{S}]|{\rm in}\rangle_{\rm dr} \approx \frac{\kappa}{4\pi}\sum_{i = 1}^n S^{(s)}_{-,i}(z) & {}_{\rm dr}\langle {\rm out}| \mathcal{S}  |{\rm in} \rangle_{\rm dr}
\ee
are expected to hold at tree-level \cite{Gabai:2016kuf, Choi:2019rlz} irrespective of whether the incoming/outgoing states are dressed. Similar relations hold for $\mathscr{M}_+$ insertions.

Combining this with  the dressing equalities \eqref{dressingid} implies that we have the identities
\be 
\label{LGT-dressed}
 \langle {\rm out}| \mathscr{M}_-(s,z) \prod_{i \in {\rm out}} \mathcal{D}(p_{i}) \mathcal{S} \prod_{j \in {\rm in}}\mathcal{D}^{\dagger}(p_{j})| {\rm in}\rangle \approx 0.
\ee
This generalizes to  $s\neq 0$  that  the S-matrix elements of soft gravitons between dressed states of the kind in \eqref{LGT-dressed} vanish \cite{Gabai:2016kuf, Choi:2019rlz}.

 \subsection{Goldstone 2-point function}

As we have seen in Section \ref{sec:Gold} the Goldstone two-point function takes the form
\be 
\label{Godstone-two-pt}
\langle 0| \mathscr{S}_{\pm}(s, z) \mathscr{S}_{\pm}^{\dagger}(s',w) |0\rangle  = \pi \kappa^2 R_{ss'}  \delta^{(2)}(z, w).
\ee
This implies the following two-point function of dressing operators 
\be 
\label{two-point}
\begin{split}
\langle p_i| \mathcal{D}_{\pm}(p_i) \mathcal{D}_{\pm}^{\dagger}(p_j) |p_j \rangle &= \exp\left\{\sum_{s,s' = 0}^{\infty} (-1)^{s}\frac{R_{ss'}}{\pi \kappa^2} \frac{\kappa^2}{( 4\pi)^2} \int \rd^2w S^{(s)}_{-,i}(w) S^{(s')}_{+,j}(w) \right\} \langle p_i|p_j\rangle\\
&= \exp\left\{\sum_{\ell  =0}^{\infty}\sum_{\ell'  =0}^{\infty} C_{ij}^{\ell \ell'} [z_{ij}^{\ell} \bz_{ji}^{\ell' } |z_{ij}|^2\log\left(H_{\ell\ell'}|z_{ij}|^2\right)] \p_{z_i}^\ell \p_{\bz_j}^{\ell'} \right\}\langle p_i| p_j \rangle := \mathcal{S}_{\rm soft}^{ij} \langle p_i|p_j\rangle,
\end{split}
\ee
where $z_{ij}:=z_i-z_j$,  $H_{\ell\ell'}$ is a constant defined in \eqref{defA}, the $i = j$ terms vanish due to the vanishing of the exponent in the second line of \eqref{two-point}, and 
\be \la{Cell}
C_{ij}^{\ell \ell'} = \frac{ \kappa^2}{2 (4\pi)^2 } \frac{1}{ \ell!\ell'!}\left(\sum_{s,s'\geq \ell, \ell'} (-1)^{s} R_{ss'}\frac{(2h_{i})_{s - \ell}}{(s-\ell)!}\frac{(2\bar{h}_{j})_{s' - \ell'}}{(s'- \ell')!}(-\eta_i \omega_i)^{1-s}(-\eta_j \omega_j)^{1-s'}\right).
\ee
We have used \eqref{Sdef} together with 
\be \la{Iij}
I_{ij}^{\ell \ell'}(z_i,\bar{z}_i) := \int \rd^2w \frac{( z_i-w)^\ell ( \bz_j-\bw)^{\ell'}}{( \bz_i-\bw)( z_j-w)} =  -2\pi (z_i - z_j)^{\ell} (\bz_j - \bz_i)^{\ell'} \log \left( H_{\ell\ell'}|z_i - z_j|^2 \right),
\ee
which we prove in Appendix \ref{app:dress}. 
Note that $I_{ij}^{\ell \ell'}$ is a positive definite operator, namely \\
$\sum_{\ell,\ell'}^{i,j} \phi^{i*}_{\ell} I_{ij}^{\ell \ell'} \phi^j_{\ell'} \geq 0$ and that
\be 
\label{eq:useful-id}
\p_{z_i}^{\ell + 1} \p_{\bz_j}^{\ell' + 1} I_{ij}^{\ell \ell'} = (2\pi)^2  \ell!\ell'!  \delta^{(2)}(z_{ij}).
\ee
The expression for $C^{\ell\ell'}_{ij}$ is a priori divergent given that $R_{ss'}$ is singular.
However it can be resummed, quite remarkably.
One first uses  \eqref{Stwo-pt}, according to which $R_{ss'}=\int_{0}^{\infty} \rd \omega \omega^{s+s'-1} e^{-\epsilon \omega}$. We can then  exchange the sum over $s,s'$ with the integral, and use the binomial identity
\be
\sum_{s\geq \ell} \frac{(2h)_{s-\ell}}{(s-\ell)!} x^s = x^\ell (1+ x)^{2h},
\ee 
which allows to sum \eqref{Cell} into 
\be
C_{ij}^{\ell\ell'} = -\frac{ \kappa^2}{2 (4\pi)^2 } \frac{(\eta_i \omega_i)^{1-\ell}(-\eta_j \omega_j)^{1-\ell'}}{ \ell! \ell'!}
\int_{0}^\infty \rd \omega e^{-\epsilon \omega} \omega^{\ell+\ell'-1}\left(1+ \frac{\omega}{\eta_i\omega_i}\right)^{2h_i}\left(1- \frac{\omega}{\eta_j\omega_j}\right)^{2\bar{h}_j}.
\ee

\begin{figure}[h!]
\includegraphics[scale = 0.26]{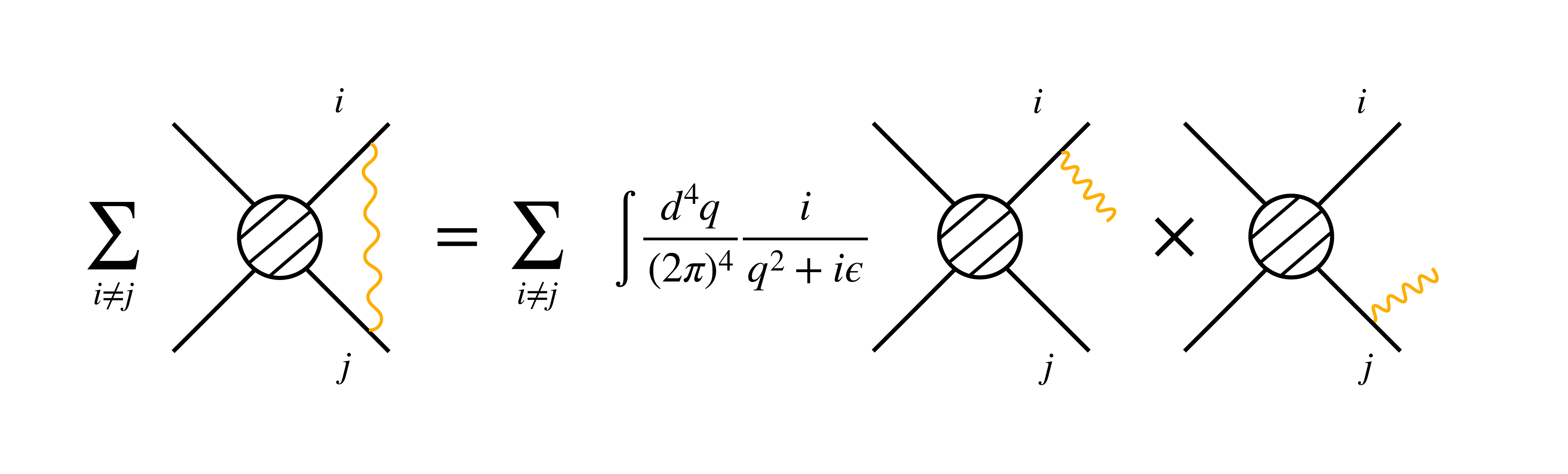}
\captionsetup{width=0.9\textwidth}
\caption{Decomposition of a ladder diagram into off-shell graviton emissions. The emissions become on-shell upon taking the real part. At tree-level, each resulting emission vertex can be approximated by a sum over soft factors.}
\label{Fig1}
\end{figure}

It is instructive to compare this with the result of exponentiating virtual graviton exchanges. The calculation by Weinberg \cite{Weinberg:1965nx} can be generalized to include the contribution from the whole tower of soft theorems to the eikonal vertices, as illustrated in Figure \ref{Fig1}. 
As a result, the ``soft'' S-matrix becomes\footnote{We have used the completeness relation for the polarization tensors $\sum_{\alpha, \beta, \gamma, \delta} \pi^{\alpha\gamma;\beta\delta} \varepsilon_{\alpha\gamma}^{\mu} \varepsilon_{\beta \delta}^{\nu} = \eta^{\mu\nu}$ to recast the integrand
into a product of soft factors.} 
\be \label{SoftW}
\mathcal{S}_{\rm soft} = \prod_{i \neq j} \mathcal{S}^{ij}_{\rm soft} = \lim_{\epsilon\to 0} \exp\left\{-\frac{2 \pi}{2 (2\pi)^4}\sum_{i,j}\sum_{s = 0}^{\infty} \sum_{s' = 0}^{\infty} \int \frac{\omega^2 \rd\omega \rd^2z}{\omega} e^{-\epsilon \omega} \omega^{s + s' - 2}S^{(s)}_{-,i}(z)  (-1)^{s'}S^{(s')}_{+,j}(z) \right\}.
\ee
The integral is a positive definite operator. 
Here, the $(-1)^{s'}$ sign in the exponent comes from the fact that the soft photon is incoming for one leg and outgoing for the other.  For $s = 0$, this result agrees with that computed by Weinberg in \cite{Weinberg:1965nx}. Importantly, for $s = 0$, the integral in the exponent is positive definite and IR divergent thereby setting all matrix elements to zero. In fact, the Hermiticity property \eqref{eq:hermiticity} together with the vanishing of \eqref{Iij} for $i = j$ non-trivially implies that the integral remains positive definite upon including all $s$ contributions.  
Performing the integration over $\omega$ reproduces the kernel $R_{ss'}$ and gives
\be \label{SoftW2}
\mathcal{S}_{\rm soft}  = \prod_{i \neq j} \exp\left\{-\frac{1}{2 (2\pi)^3}\sum_{s = 0}^{\infty} \int  \rd^2z (-1)^s\widehat{S}^{(s)}_{-,i}(z) S^{(s)}_{+,j}(z) \right\},
\ee
where we have introduced the  Goldstone mode operator
\be \label{eq:Gsum}
\widehat{S}^{(s)}_{\pm, i} (z):=  \sum_{s'} R_{ss'} S^{(s')}_{\pm, i}(z).
\ee 
This relation is analogous to \eqref{relSM} and  provides the action of the Goldstone charge $\mathscr{G}_\pm (s,z)$ on the external states.

It is expected but still remarkable that this evaluation can be reproduced as the expectation value of dressing operators 
\be 
\langle {\rm out}| {\prod_{i}}\mathcal{D}(p_i)  |{\rm in}\rangle = \mathcal{S}_{\rm soft}.
\ee
To derive this result we use the formula \eqref{eq:toy-correlator} in Appendix \ref{sec:state-factorization}. The $i = j$ contributions vanish as explained after \eqref{two-point}. In \eqref{SoftW2} the would-be $i = j$ terms correspond to diagrams where the internal graviton attaches to the same external leg (see Figure \ref{Fig1}).

If one restricts  the dressing \eqref{eq:Dress} to include only the $s = 0$ term, one then reproduces the conformal primary (Lorentz invariant) Faddeev--Kulish dressings introduced in 
\cite{Himwich:2020rro, Arkani-Hamed:2020gyp}. These vertex operators were shown in \cite{ Himwich:2020rro} to compute the infrared soft S-matrix in gravity. 
\eqref{eq:Dress} on the other hand represents the gauge invariant generalization of the conformal primary dressings in \cite{Arkani-Hamed:2020gyp} resulting from the infinity of conservation laws associated with the tower of tree-level soft theorems.  

Note that \eqref{Godstone-two-pt}, together with \eqref{eq:useful-id}, imply that 
\be 
\label{Gtwo-pt}
\langle 0|\mathscr{G}_{\pm}(s, z) \mathscr{G}_{\pm}^{\dagger}(s', z')|0\rangle =  \frac{\kappa^2 R_{ss'}}{2 s! s'!}(z - z')^s(\bz'-\bz)^{s'}|z-z'|^2 \log\left(H_{ss'}|z - z'|^2\right).
\ee
In particular, for $s = 0$ this precisely agrees with the two point function of Goldstone operators derived in \cite{Himwich:2020rro}.\footnote{Note that our normalization for the Goldstone operator differs  by a factor of $\frac{i}{4\pi}$ from that in \cite{Himwich:2020rro}, namely $-\langle p| \frac{i}{4\pi}\left(\mathscr{G}_{+}(0) + \mathscr{G}_{-}(0) \right)= \langle p| C_{}$. Moreover, recall that in contrast to \cite{Himwich:2020rro} we use a parameterization \eqref{qhat} for the momenta  which is such that $p_i \cdot p_j = \epsilon_i \epsilon_j \omega_i \omega_j |z_{ij}|^2$. } 
One puzzle that we are facing is that the  Goldstone 2-point function \eqref{Gtwo-pt} is increasingly divergent when $s\geq 1$ since $R_{ss'}\sim \epsilon^{-(s+s')}$.
 Moreover this divergence is for $s\geq 1$ a UV divergence since $\epsilon\to 0$  appears in the integrals as a UV regulator $ e^{-\epsilon \omega}$. One is left to wonder whether it is possible to understand these UV divergent correlators in a more regular manner? This is what we investigate next where we show that they can be understood in terms of Goldstone insertions.


\subsection{The Goldstone insertion}

 In the previous section we have seen that the soft S-matrix involves the insertion of a Goldstone  operator 
 $\widehat{S}^{(s)}_\pm(z,p)=\sum_{s'}R_{ss'} {S}^{(s')}_\pm(z,p)$.
 This expression is formal and doubly divergent since the sum involves negative powers of $\epsilon$, as well as growing factorial $\Gamma(s+s')$. However, it turns out that we can resum this formula into a meaningful expression.

 This follows from the analysis done in Sections \ref{sec:Ramanujan} and \ref{sec:spot}. There we have seen that 
 the relationship $\widehat{S}^{(s)}_\pm(z,p)=\sum_{s'} R_{ss'} {S}^{(s')}_\pm(z,p)$ should be interpreted as the relationship between the evaluation of a Mellin space operator  $S_\pm(\Delta;z,p)$ at positive and negative integer dimensions, namely
 \be  
 \mathrm{Res}_{\Delta = 1-s}[ S_\pm(\Delta;z,p)] = S_\pm^{(s)}(z,p),\qquad 
 S_\pm(\Delta=1+s;z,p)= \widehat{S}^{s}_\pm(z,p).
 \ee 
 The celestial OPE analysis has revealed such an operator \cite{Pate:2019lpp,Guevara:2021abz,Himwich:2021dau,Guevara:2021tvr,Jiang:2021ovh}. It is given, when $p_i=\eta_i\omega_iq(z_i)$, by 
\be
S_{-}(\Delta;z,p_i) &=
\frac{\kappa}{ 2}\,  ({\eta_i}\omega_{i})^{\Delta}
\left(\frac{z-z_i}{\bz-\bz_i}\right) 
\sum_{\ell = 0}^{\infty} \frac{\Gamma(\Delta - 1 + \ell)\Gamma( 2h_i + 1)}{\ell! \Gamma(\Delta +2h_i + \ell)}
 (z-z_i)^{\ell}\p_{z_i}^\ell ,
\ee
where $2h_i=-\omega_i\pa_{\omega_i} + s_i$.  $S_{+}(\Delta;z,p_i)$ is given by the replacements $z\leftrightarrow\bz$ and $2h_i \to 2\bar h_i$ (see \eqref{hi}).
Using that $\mathrm{Res}_{\Delta = -n} \Gamma(\Delta)= \frac{(-1)^n}{n!} $, we get that the residue of this operator at  $\Delta=1-s$ is  
\be 
\mathrm{Res}_{\Delta = 1-s} [S_-(\Delta;z,p_i)] 
&= \frac{\kappa}{ 2}\, ({\eta_i} \omega_i)^{1-s}
\left(\frac{z-z_i}{\bz-\bz_i}\right)\sum_{\ell = 0}^s \frac{(-1)^{\ell-s}}{\ell!} \frac{(2h_{i})_{s - \ell}}{(s-\ell)!}   (z - z_i)^{\ell} \p^\ell_{z_i},
\ee 
which is equal to $S_-^{(s)}(z,p)$ (see \eqref{SoftT}).
Evaluating this operator at $\Delta = 1+s $ instead, leads to our proposal for the resummed Goldstone operator
\be
\label{eq:ansatz}
\widehat{S}^{(s)}_-(z,p_i) &=
\frac{\kappa}{ 2}\,  ({\eta_i}\omega_{i})^{1+s}
\left(\frac{z-z_i}{\bz-\bz_i}\right) 
\sum_{\ell = 0}^{\infty} \frac{\Gamma(s+ \ell)\Gamma( 2h_i + 1)}{\ell! \Gamma(2h_i +1+ s+\ell)}
 (z-z_i)^{\ell}\p_{z_i}^\ell.
\ee
This operator is our ansatz for the resummation of \eqref{eq:Gsum}. It is well defined if $s>0$ while for $s=0$ we should omit the term $\ell=0$ in the sum.  We have not attempted to prove this formula by explicitly resumming \eqref{eq:Gsum}, so \eqref{eq:ansatz} should be regarded as a physically motivated guess. We leave a proof of this to future work. 
This Ansatz means  that we expect the dressing operator \eqref{eq:Dress} to be equal to 
\be 
\label{eq:Dress2}
     \mathcal{D}(p) &=  \exp \left\{\sum_{s = 0}^{\infty} \frac{1}{(2\pi)^2 \kappa}\int d^2z \left[(-1)^{s}\widehat{S}_-^{(s)}(z;p)\mathscr{M}(s,z) - \widehat{S}_+^{(s)}(z;p)\mathscr{M}^\dagger(s,z) \right] \right\},\cr 
     &= \exp \left\{\sum_{s = 0}^{\infty} \frac{1}{(2\pi)^2 \kappa}\int d^2z \left[(-1)^{s}{S}_-^{(s)}(z;p)\mathscr{S}(s,z) - {S}_+^{(s)}(z;p)\mathscr{S}^\dagger(s,z) \right] \right\}.
\ee
The expression \eqref{SoftW2} for the soft S-matrix then trivially follows from these expressions and the commutators 
\be\la{corn-comm2}
[\mathscr{M}(n,z ),\mathscr{S}^\dagger(m, z')]= \pi \k   \delta_{n,m}\d^2(z,z').
\ee

\section{Higher spin charges}\la{sec:charges}

In this section we write the tower of charges introduced in \cite{Freidel:2021qpz, Freidel:2021dfs, Freidel:2021ytz} as corner charges expressed in terms of the memory and Goldstone variables.

The result we establish in this section can be summarized as follows: We first use that the integrated charges can be decomposed as a sum $Q_s= Q_s^1 + Q_s^2 +\cdots$ of soft plus hard plus higher order contributions. We then show that the higher spin charges can be decomposed  into positive and negative helicity sector as
\be
 Q_s(\tau) =
  -\frac1{4}\left[ i^s Q_-(s)  + i^{-s}Q_+^*(s)\right],
\ee
and that each contribution can be
written as  a corner charge in terms of $\mathscr{M}(n,z), \mathscr{S}(n,z)$. 
 For the soft charges we have seen that $Q^{1*}_{s+}(\tau)
 =  \int_S   \tau(z) D^{s+2}\mathscr{M}^*_+(s,z)$ while for the  quadratic charges, their corner expression  read \footnote{In this section we use the shorthand $D=D_z$ and $\bar{D}=D_{\bz}$}
\be
Q^{2*}_{s+}(\tau)
 &=-\frac{  1}{2\pi}  \sum_{n=0}^\infty \sum_{\ell=0}^s (-)^{\ell+s} (\ell+1)
\f{(s+n-\ell)_{s-\ell}}{(s-\ell)!}
 \int_S   D^{s-\ell}\tau_s(z) \mathscr{S}_+(n,z)D^{\ell}  {\mathscr{M}}^*_+(s+n-1,z).
 \ee

\subsection{Complex mass aspect}

We start with the complex mass aspect or the spin-0 charge. This is given by \cite{ Freidel:2021dfs}
\be\la{mC}
m_\C(z)=  
\int_{-\infty}^{+\infty} \rd u \left(\frac12 D^2 N(u,z) +\frac 14 C(u,z) \dot{N}(u,z)\right).
\ee
For the linear/soft term we simply have
\be 
\frac12 \int_{-\infty}^{+\infty} \rd u D^2 N(u,z)= \frac1{4} D^2 \left[\mathscr{M}_-(0,z)+ \mathscr{M}_+^\dagger(0,z)\right].
\ee
The quadratic term reduces to a sum of positive and negative helicity contributions
\be
\int_{-\infty}^{+\infty} \rd u \, C(u,z) \dot{N}(u,z)
&= \int_{-\infty}^{+\infty} \rd u \, \left(C_+(u,z)\dot{N}_+^\dagger(u,z) + C_-^\dagger(u,z) \dot{N}_-(u,z) \right).
\ee
Using the expansion \eqref{uexp} we then find
\be
\int_{-\infty}^{+\infty} \rd u \, C_+(u,z)\dot{N}_+^\dagger(u,z)
&=
- \frac{1}{2 i\pi} \int_{-\infty}^{+\infty} \rd u
\left( \sum_{n=0}^\infty \frac{u^n}{n!}( i^{-n} \mathscr{S}_+(n,z))\right) \dot{N}_+^\dagger(u,z)\cr
  &=-\frac{1}{2\pi} \sum_{n=1}^\infty 
   \mathscr{S}_+(n,z)  
  \left(i^{n-1} \int_{-\infty}^{+\infty} \rd u \frac{u^{n-1}}{(n-1)!}  N_+(u,z)\right)^\dagger\cr
  &=-\frac{1}{4\pi} \sum_{n=1}^\infty 
 \mathscr{S}_+(n,z)  \mathscr{M}_+^\dagger(n-1,z). 
\ee
In the third equality we integrated by parts and used the boundary conditions $[u^n N_+(u,z)]_{-\infty}^{+\infty}=0.$
Similarly we find
\be
\int_{-\infty}^{+\infty} \rd u \, C_-^\dagger(u,z)\dot{N}_-(u,z)= -\frac{1}{4\pi} \sum_{n=1}^\infty 
 \mathscr{S}_-^\dagger(n,z) \mathscr{M}_-(n-1,z).
\ee
Putting everything together, the mass aspect is $\cM_\C=\frac1{4}(\cM_- +\cM_+^\dagger),$ where
\be
\cM_+^\dagger(z) &=   D^2 \mathscr{M}_+^\dagger(0,z)  - \frac{1}{4\pi} \sum_{n=0}^\infty  
 \mathscr{S}_+(n+1,z) \mathscr{M}_+^\dagger (n,z),
 \cr
 \cM_-(z) &=   D^2 \mathscr{M}_-(0,z)  - \frac{1}{4\pi} \sum_{n=0}^\infty  
 \mathscr{S}^\dagger_-(n+1,z) \mathscr{M}_- (n,z).
\ee

\subsection{Renormalized higher spin charges}\la{sec:renq}

As shown in \cite{Freidel:2021dfs}, the mass charge aspect \eqref{mC} has a finite action on the gravitational phase space at null infinity; however, for charges of spin $s\geq 1$ a renormalization procedure is required.
 The renormalized higher spin generators are defined as \cite{Freidel:2021dfs, Freidel:2021ytz}
 \be\la{Qren}
 \hat q_s(u,z):=  \sum_{n=0}^{s} \frac{(-u)^{s-n}}{(s-n)!} D^{s-n} \cQ_n(u,z)\,,
\ee
where $\cQ_n(u,z)$ are solutions to the recursion relation
\be\la{highQ}
\cQ_{n} &= D\pa_u^{-1} (\cQ_{n-1}) + \frac{(n+1)}{2} \pa_u^{-1}(C \cQ_{n-2})\,.
\ee
From \eqref{Qren} the renormalized higher spin charge aspects are obtained as the limit
\be\la{Qren2}
q_s(z)=\lim_{u\to-\infty} \hat q_s(u,z).
\ee 
By smearing these charge aspects with arbitrary spin-$s$ transformation parameters, we obtain the renormalized charges
\be\la{Qstau}
Q_s(\tau) := \int_S  q_s(z) \tau_s(z).
\ee
We now want to express these renormalized charges in terms of the variables parametrizing the corner symplectic potential \eqref{Directe}.
The charge aspects solving \eqref{highQ} admit an expansion in terms of powers of radiation fields as
\be\cQ_s =\sum_{k=1}^{\max[2,s+1]} \cQ_s^{k}\,,
\la{charge-kexp}
\ee
where 
$\cQ_{-2}= \frac12 \dot{N}$, $\cQ_{-1}= \frac12 D N$.
Moreover, each $\cQ_s^k$ obeys the following recursion relation
\be
\la{charge-k-rec}
\cQ_{s}^{k} &= D\pa_u^{-1} (\cQ_{s-1}^{k}) + \frac{(s+1)}{2} \pa_u^{-1}(C \cQ_{s-2}^{k-1}).
\ee
From \eqref{charge-k-rec} one can derive the relation of the charges at order $k$ in terms of those at order $k-1$, for $k\geq 1$, as
\be\la{Qk}
\cQ^k_{n}(u,z)&= \frac12
\sum_{\ell=0}^{n} (\ell+1) \p_u^{-1}(\p_u^{-1} D)^{n-\ell} \left[C (u,z) \cQ^{k-1}_{\ell-2}(u,z)  \right].
\ee
When $k\geq 3$ the sum over $\ell$ can be restricted to be for $\ell \geq k$ as the first terms vanish. 
This can be used to write the renormalized aspects\eqref{Qren} as 
\be\la{qren}
\hat q^k_s(u,z)&=\sum_{n=k-1}^{s} \frac{(-u)^{s-n}}{(s-n)!} D^{s-n} \cQ^k_n(u,z)\cr
&=\frac12 \sum_{n=k-1}^{s}\sum_{\ell=0}^{n}  (\ell+1)  \frac{(-u)^{s-n}}{(s-n)!} 
(\p_u^{-1} )^{n-\ell+1}D^{s-\ell}  \left[C (u,z) \cQ^{k-1}_{\ell-2}(u,z)  \right]\cr
&=\frac12 \p_u^{-1} \sum_{\ell=0}^{s}  (\ell+1)  \frac{(-u)^{s-\ell}}{(s-\ell)!} 
 D^{s-\ell}  \left[C (u,z) \cQ^{k-1}_{\ell-2}(u,z)  \right]\,,
 \ee
where in the last passage we have switched sums and used the generalized Leibniz rule 
\be\la{Leibn}
\pa_{u}^{-1}\left( \frac{(-u)^k}{k!} f(u)\right)
&=\sum_{n=0}^{k}\frac{(-u)^{(k-n)}}{(k-n)!}
\pa_{u}^{-(n+1)}f(u).
\ee
In the following sections we derive expressions for these higher spin charges for arbitrary $s$ and $k = 1, 2.$ For $k\geq3$ we derive a recursive formula (the explicit expression of the cubic charges $k=3$ in the new discrete basis is derived in \cite{Freidel:2023gue}).

\subsection{Positive and Negative helicity charges}

Let us emphasize that the decomposition into positive and negative helicity sector defined for the soft sector in \eqref{soft-charges-0} can be extended to all the non-linear charges. This means that we can define 
\be
 Q_s(\tau) =
  -\frac1{4}\left[ i^s Q_{s-}(\tau)  + i^{-s}Q_{s+}^\dagger(\tau)\right].
\ee
As we have in \eqref{qsoftC}, this means for the linear charges that 
\be
 Q^1_{s-}(\tau)
 = (-1)^s \int_S   D^{s+2}\tau(z) \mathscr{M}_-(s,z),\qquad 
 Q^1_{s+}(\tau)
 = (-1)^s \int_S   \bar{D}^{s+2}\tau(z) \mathscr{M}_+(s,z)\,.\la{Q1}
\ee

The quadratic contribution to the renormalized higher spin charge aspects \eqref{Qren2} is given by (see Appendix \ref{App:Q2} for details)
\be
 q^2_s(z)&:= \lim_{u\to-\infty} \hat q^2_s(u,z)
\ee
with
\be
\hat q^2_s(u,z)
&=\frac14\p_u^{-1}  \left[\sum_{\ell=0}^s \frac{(\ell+1) (-u)^{s-\ell}}{(s-\ell)!} 
  D^{s-\ell}\left[C  D^{\ell} \p_u^{-\ell+1}  N  \right](u,z) \right]\,.
\ee

This expression of the charge aspects allows us to write  the higher spin quadratic charges as the single integral of a charge density as 
\be
Q^2_s(\tau)  &= {-}\frac14 
\sum_{\ell=0}^s \frac{(\ell+1) }{(s-\ell)!}  \int_{\scri} 
  C(u,z) \left[D^{\ell} (\hD -1)_{s-\ell} \p_u^{1-s}  N(u,z) \right]    D^{s-\ell} \tau_s(z),\cr
  & = {-} \frac14 \sum_{\ell=0}^{s} \frac{(\ell+1)}{(s-\ell)!}
    \int_{\scri}    {C^\dagger} (u,z)    \left[D^{\ell} (\hD +3)_{s-\ell} \p_u^{1-s}{N^\dagger}(u,z) \right]
    D^{s-\ell} \tau_s(z)\,,\la{Qs}
\ee
where we have introduced the operator $\widehat\Delta :=\pa_u u$, and in general
\be
Q_s^k(\tau) := \int_S  q_s^k(z) \tau_s(z)
\ee
is the integrated charge operator and we have denoted $\int_{\scri}:= \int_{-\infty}^{+\infty} \rd u \int_S \rd^2 z \sqrt{q}$. 

From the two expressions \eqref{Qs}, it is straightforward to verify that  the action of the quadratic higher spin charges on the conformal graviton is given by \cite{Freidel:2021ytz}
\be
[ Q^2_s(\tau), {N}_\pm(\Delta)]=\frac{\kappa^2}{8} \sum_{\ell=0}^s
 (\ell+1 )\frac{(\Delta +1 \pm 2)_{s-\ell}}{(s-\ell)!} D^{s-\ell} \tau_s D^{\ell}{N}_\pm({\Delta+1  -s}) \,.
\ee

As said above, the quadratic charge is expressible as a sum 
\be\label{Qchargesplit}
 Q^2_s(\tau) =
  -\frac1{4}\left[ i^s Q^2_{s-}(\tau)  + i^{-s}Q_{s+}^{\dagger2}(\tau)\right].
\ee
In terms of the  soft and Goldstone variables
$ \mathscr{M}(n), \mathscr{S}(n)$, the + component reads
\be
Q^{2 \dagger}_{s+}(\tau)
 &=- \frac{  1}{4\pi}  \sum_{n=0}^\infty \sum_{\ell=0}^s (-)^{\ell+s} (\ell+1)
\f{(s+n-\ell)_{s-\ell}}{(s-\ell)!}
 \int_S   D^{s-\ell}\tau_s(z) \mathscr{S}_+(n,z)D^{\ell}  {\mathscr{M}}^\dagger_+(s+n-1,z) 
 \cr
&=-\frac{(-)^s}{ 4\pi}\sum_{n=0}^\infty \sum_{\ell=0}^{s} (\ell+1) \frac{ (3-n)_{s-\ell} }{(s-\ell)!}
    \int_S  D^{s-\ell} \tau_s(z)
     \mathscr{S}^\dagger_+(n,z)  D^{\ell} \mathscr{M}_+(s+n-1,z)  
 \,,\la{Q2}
\ee
where the two expressions follow respectively from the first and second line of \eqref{Qs} (again see  Appendix \ref{App:Q2} for a detailed derivation).
The same expression holds for   $Q^2_{s-}(\tau)$ in terms of the negative components of the corner variables.

The action of the quadratic charge \eqref{Q2} on the Goldstone operators is given by (see Appendix \eqref{App:Q2})
\be\la{Q2Nn}
[ Q^{2\dagger}_{s,+}(\tau), \mathscr{S}_+(n,z) ]&=-
\frac{(-)^{s} \k}{ 4}    \sum_{k=0}^s 
 (s-k+1 ) 
 \left(\begin{matrix}
n+3\\
k
\end{matrix}\right)
  D^{k} \tau_s(z)
  D^{s-k}\mathscr{S}_+(n-s+1,z)\,. 
\ee

\subsection{Order $k>2$ charges}

For the order $k>2$ contributions to the charge aspects \eqref{Qren2} we derive a recursion relation.
We start with the general expression for the non-renormalized charge aspects solving \eqref{charge-k-rec}. This is given by
\be
\cQ^k_s(u,z)&= \frac12
\sum_{\ell=k}^{s} (\ell+1) \p_u^{-1}(\p_u^{-1} D)^{s-\ell} \left[C (u,z) \cQ^{k-1}_{\ell-2}(u,z)  \right].
\ee

As shown in Appendix \ref{App:Qk}, the renormalized charge aspects at a given order $k$ can be written in terms of the aspects at order $k-1$ as
\be
\hat q^k_s(u,z)&=\sum_{n=k-1}^{s} \frac{(-u)^{s-n}}{(s-n)!} D^{s-n} \cQ^k_n(u,z)\cr
&=\frac12 \p_u^{-1}\sum_{m=0}^{s-k} \f{(-u)^m}{m!} D^{m} C (u,z)
\left[(s-m+1) \hat q^{k-1}_{s-m-2}(u,z)+
uD\hat q^{k-1}_{s-m-3}(u,z)
\right]\,.
\ee

Therefore, the $k$-th order contribution to the higher spin charges reads
\be
Q^k_s(\tau)=-\frac12 \sum_{m=0}^{s-k} \int_{\scri}\f{(-u)^m}{m!} \tau_s(z)D^{m} C (u,z)
\left[(s-m+1) \hat q^{k-1}_{s-m-2}(u,z)+
uD\hat q^{k-1}_{s-m-3}(u,z)
\right]\,.
\ee
This formula, together with \eqref{uexp}, \eqref{Q1} and \eqref{Q2}, allows for the recursive generation of the nonlinear contributions to the higher spin charges in terms of the memory and Goldstone variables. We leave the implications of this formula for the calculation of the non-linear charge algebra to future analysis.

\section{Conclusions}\la{sec:conc}

We have shown in this work that, for radiation signals that belong to a Schwartz space where both an IR (exponential decay for $u\to\pm \infty$) and a UV (exponential decay for $\omega\to +\infty$) completeness conditions are satisfied, one can introduce a discrete basis for the radiative phase space at null infinity. This basis is formed by the memory observables and the conjugate Goldstone gravitons defined respectively in \eqref{Mn} and \eqref{Goldstones}, and it is labelled by a non-negative integer number. This allowed us to recast the radiative phase space at $\scri$ as a corner phase space on an arbitrary cut.

We have exploited this new discrete basis in three applications, providing a consistency test of our analysis. 
First, we have used the Goldstone operators to define generalized dressed states  \eqref{eq:Dress} that diagonalize the whole tower of higher spin soft charges. These states were used to construct the soft part of gravitational scattering amplitudes at tree-level, where the infinity of charge conservation laws is now implemented. We also have constructed the higher spin generalization of the Goldstone two-point function.
This provides a fully gauge invariant generalization of the leading order ($s=0$) \cite{Himwich:2020rro, Arkani-Hamed:2020gyp} and sub-leading order ($s=1$) \cite{Pasterski:2021dqe} construction of the soft S-matrix in terms of correlation functions of exponentiated Goldstone operators. 
Second, we have shown that the Goldstone dressing can also be expressed in terms of the discrete memory observables and revealed a duality between the soft graviton and the Goldstone insertions in the decomposition of the ladder diagram into soft emissions. 
Third, we have shown how the infinite tower of celestial charges constructed in \cite{Freidel:2021ytz} can be expressed,  both at the linear and the quadratic orders, in terms of the memory and
Goldstone variables. This way, we have recast 
all the higher spin charges as corner integrals, removing the integral over the  $\scri$ time direction.

Beyond these initial applications, introducing a discrete basis for conformal primary wavefunctions presents several advantages. From a more conceptual point of view, it naturally resolves the tension between the original basis given by the principal continuous series $\Delta \in 1+i\R$ obtained in \cite{Pasterski:2017kqt}  and the derivation of the soft theorems from the OPE of two conformal primary gravitons which involves the conformally soft limit $\Delta\to 1-s$. An argument for understanding the insertion of this conformally soft
graviton as an analytical continuation of conformal dimensions through certain contour integrals along
the principal continuous series was presented in \cite{ Donnay:2020guq}. The results of this work provide a  uniform and more transparent treatment. From a more technical perspective, the computational advantages of the new discrete basis can help investigate the mixed helicity sector of the charge algebra to elucidate the fate of the $w_{1+\infty}$ structure in full general relativity.

A related application regards the formulation of a dictionary between the 
multipole moments of the gravitational radiation emitted by a source and the complex charges forming the higher spin algebra initiated in \cite{Blanchet:2020ngx, Compere:2022zdz}. Extension beyond the linearized theory would represent an exciting step in the programme of reconstructing an asymptotically flat metric from holographic data.

In our analysis, we have proposed, through \eqref{eq:Gsum} and \eqref{eq:Dress2}, that the higher spin soft graviton insertions can be re-summed as the insertion of a higher spin Goldstone operator. Part of the derivation involved in this resummation is conjectural and deserves a more thorough mathematical study. Nevertheless, it opens up the fascinating possibility to recast the gravitons amplitudes entirely in terms of the insertions of higher spin memory or Goldstone operators. We hope to come back to this question in the future.

It would be extremely interesting to understand precisely to what extent such correlation functions approximate the infrared-finite gravitational scattering amplitude and whether similar ideas can be used to systematically calculate scattering amplitudes beyond the leading eikonal regime. 
Another interesting aspect to investigate is whether celestial amplitudes involving Goldstone insertions obey special identities paralleling the Ward identities for tree-level soft insertions. Our expansion for the tower of charges in terms of Goldstone and conformally soft operators suggests this may be the case. 
 Such an identity has been used in \cite{Pasterski:2022djr, Donnay:2022hkf}
 for the  $s=0$ Goldstone operator to recast loop correction as the dressing of the charge operators. This required to derive the IR-divergent part of the corrected subleading soft graviton theorem as a Ward identity for a certain ``symmetry-improved'' subleading soft charge. It would be remarkable to establish from first principle this identity. And one can wonder if the dressing mechanism, including loop corrections, can be generalized to the whole tower of Goldstone insertions using the techniques we develop here.
 Eventually, one may ask if our all-order dressings can also play a role in extending the description of the manifold of CCFT vacua initiated in \cite{Kapec:2022hih} to subleading orders in the soft expansion.

We conclude with a word of caution. 
It is well known that the soft theorems receive loop corrections \cite{Laddha:2018myi, Laddha:2018vbn, Sahoo:2018lxl}. These corrections introduce log terms in the soft expansion 
\be 
\widetilde{N}_\pm(\omega)
= \sum_{n=0}^\infty \sum_{m=0}^n \omega^n (\log \omega)^m \mathscr{M}_\pm(n,m),
\ee 
where the soft order bounds the power of the log terms: $\omega^n (\log \omega)^m$ with $m\leq n$. It implies that the Mellin transform admits poles of higher order in the external dimensions. More precisely the poles at $\Delta =1-n$ are of at most order $n$:
\be \la{p-exp}
 \widehat{N}_\pm (\Delta) \asymp \sum_{n}^{\infty} 
 \sum_{m=0}^n\frac{\mathscr{M}_\pm(n,m)}{(\Delta + n)^m},
 \ee
 where $\asymp$ denotes the asymptotic pole expansion.
 For such states, we cannot argue any longer that the higher spin charges are well defined, and we leave the analysis beyond the single pole states to future investigation. 
 
 Finally, our analysis only addresses the evaluation of the real part of the higher spin soft graviton exchange. For completeness, we would have to investigate if one can access the virtual imaginary component of that exchange in a similar manner.

\section*{Acknowledgements}

We are grateful to Jan de Boer, Luca Ciambelli, Ben Freivogel, Dan Kapec,  Kevin Nguyen and Sabrina Pasterski for discussions. 
A.R. presented our  work in the ``Quantum gravity around the corner'' conference organized by L. Ciambelli and C. Zwikel at Perimeter institute in Oct. 2022.
Research at Perimeter Institute is supported in part by the Government of Canada through the Department of Innovation, Science and Economic Development Canada and by the Province of Ontario through the Ministry of Colleges and Universities. D.P. has received funding from the European Union's Horizon 2020 research and innovation programme under the Marie Sklodowska-Curie grant agreement No. 841923. A.R. is additionally supported by the Stephen Hawking fellowship at Perimeter Institute and the Heising Simons foundation as part of the QuRIOS collaboration at the University of Amsterdam.

\appendix
\section{Properties of the Fourier and Mellin transform}
\la{App:A}

In this appendix we collect some properties of the position, momentum and Mellin space representations of the shear. It is straightforward to show using the definitions that $C(u)$, $\widetilde{C}(\omega)$, $\widehat{C}(\Delta)$ satisfy
\be 
\label{trans-prop}
\left[C(u)\to \pa_u C(u) \right] &\equiv \left[ \widetilde{C}_{\pm}(\omega)\to  -i\omega \widetilde{C}_{\pm}(\omega) \right] \equiv
\left[ \widehat{C}_{\pm}(\Delta) \to   -i\widehat{C}_{\pm}(\Delta+1)\right],\cr
\left[C(u)\to  u  C(u)\right] &\equiv \left[ \widetilde{C}_{\pm}(\omega)\to  -i\pa_\omega \widetilde{C}_{\pm}(\omega) \right] \equiv
\left[ \widehat{C}_{\pm}(\Delta) \to i (\Delta-1) \widehat{C}_{\pm}(\Delta-1)\right],\cr
\left[C(u)\to  (\pa_u u) C(u)\right] &\equiv \left[ \widetilde{C}_{\pm}(\omega)\to  -(\omega\pa_\omega) \widetilde{C}_{\pm}(\omega) \right] \equiv
\left[ \widehat{C}_{\pm}(\Delta) \to  \Delta \widehat{C}_{\pm}(\Delta)\right].
\ee
The last line shows that the Mellin transform diagonalizes the rescaling operator $\omega\pa_\omega$ with eigenvalue $-\Delta$ \cite{Pasterski:2017kqt}. 

We conclude with a derivation of the contour prescription in \eqref{softC}. We start with
\be 
 \mathscr{M}_+(n) := \frac{i^n}{n!} \left( {\oint}_U\rd u\, u^n \pa_uC_+(u) \right)
 \ee
where $U$ a contour in the upper half complex $u$ plane.  Using 
\eqref{C+-} we find
\be 
\begin{split}
 \mathscr{M}_+(n) 
 &= -\frac{i^n}{n!} \frac{1}{2\pi i} {\oint}_U\rd u\, u^n \pa_u \int_{-\infty}^{\infty} du'\frac{1}{u' - u + i\epsilon} C(u')
 \cr
 &=
 \lim_{\omega\to0+}
 -\frac{i^n}{n!} \frac{1}{2\pi i}  \int_{-\infty}^{\infty} du'\pa_{u'}  C(u')
 {\oint}_U\rd u\, 
 \frac{u^n e^{i\omega u}}{u' - u + i\epsilon}
 \cr
 &= \lim_{\omega\to0+} \frac{i^n}{n!} {\int}_{-\infty}^{\infty}\rd u'\, e^{i\omega u'}(u' + i\epsilon)^n \pa_{u'} C(u'),
 \end{split}
\ee
where  in the second equality we evaluated the contour integral after noting that $\p_u \frac{1}{u' - u + i\epsilon} = -\p_{u'} \frac{1}{u' - u + i\epsilon}$ and integrating $\p_{u'}$ by parts. Note that it is necessary to regulate the integrand  such that the contribution from the contour at infinity vanishes.

\section{Signal reconstruction from Goldstones}
\label{app:ramanujan}
In this appendix we show that the Ramanujan master theorem can be used to reconstruct the news signal from the tower of Goldstone modes $\mathscr{S}(n), n \geq 0.$ We start from the definition
\be
i\widehat{N}_+(\Delta) &  
=  \Gamma(1+\Delta ) \int_{-\infty}^{+\infty} \rd u   (-i u{+\epsilon})^{-\Delta -1 } C(u) \cr
&=\Gamma(1+\Delta )\left[ \int_{0}^{+\infty} \rd u \left[  (-i u{+\epsilon})^{-\Delta -1 } C(u)+   (i u{+\epsilon})^{-\Delta -1 } C(-u)\right] \right],
\cr
&= i\Gamma(1+\Delta )\left[ \int_{0}^{+\infty} \rd u u^{-\Delta -1} \left[   i ^{\Delta } C(u)-     i^{-\Delta  } C(-u)\right] \right]\,,
\ee
where by $i^\Delta:= e^{i\Delta\frac{\pi}{2}}$. To get this we used that the principal branch of the logarithm is such that 
\be 
\ln(-iu +\epsilon) = \ln |u| - i{ \mathrm{sign}(u)} \frac{\pi}{2}\,.
\ee 
Moreover, as explained before, the $i\epsilon$ prescription allows us to replace $C(u)$ by $C_+(u).$
This implies that 
\be
\widehat{N}_+(-\Delta)
&=
\Gamma(1-\Delta )\left[ \int_{0}^{+\infty} \rd u u^{\Delta -1} \left[  i^{-\Delta } C_+(u)-     i^{\Delta  } C_+(-u)\right] \right]\,.
\ee
Using that 
\be 
C_{\pm}(u)= \f{i}{2\pi } \sum_{n=0}^\infty \frac{(-u)^n}{n!}  [i^{n}\mathscr{S}_\pm(n)]\,,
\ee
we get, from the Ramanujan master theorem, that 
\be
\widehat{N}_+(-\Delta)&= \f{i}{2\pi } 
\Gamma(1-\Delta) \Gamma(\Delta) [i^{-2\Delta}  - i^{2\Delta} ]\mathscr{S}_+(-\Delta)\cr
&= \frac{\pi}{\sin{\pi\Delta}}
\frac{( e^{i\pi\Delta}- e^{-i\pi\Delta })}{2i\pi}\mathscr{S}_+(-\Delta)\,.
\ee
The calculation for $\widehat{N}_-$ is analogous.

\section{Symplectic potential}

\label{App:symplectic-potential}

Using that $N(u)=\pa_u C^*(u)$ and the Fourier decomposition
\be
\widetilde{C}(\omega) = \int_{-\infty}^{\infty} du e^{i\omega u} C(u), \quad \omega \in \mathbb{R},
\ee
we see that the symplectic potential takes the form
\be
\label{symplectic-form-app}
\theta  =  \f2\k \int_{-\infty}^{+\infty} \rd u N(u) \delta C(u)
={\frac1{\pi\k}} \int_0^{+\infty} \rd\omega\left[ \tN(\omega) \delta \widetilde{C}(-\omega) +  \tN(-\omega) \delta \widetilde{C}(\omega) \right].
\ee
Hence,
\be
\theta &=\frac1{\pi\k} \int_{-\infty}^{+\infty}  \rd\omega  \tN(-\omega) \delta \tC(\omega)
= - \frac1{i\pi \k} \int_{-\infty}^{+\infty}  \rd\omega  \omega \tC^*(\omega) \delta \tC(\omega)\cr
&= \frac1{i\pi \k}  \int_0^{+\infty} \rd\omega \omega\left[  \tC^*(-\omega) \delta \tC(-\omega) - \tC^*(\omega) \delta \tC(\omega) \right] \cr
&= \frac1{i\pi \k}  \int_0^{+\infty} \rd\omega \omega\left[  \tC_+(\omega) \delta \tC_+^*(\omega) - 
\tC_-^*(\omega) \delta \tC_-(\omega) \right] \cr
&= \frac1{i\pi \k} \int_0^{+\infty} \rd\omega \omega\left[  \tC_+(\omega) \delta \tC_+^*(\omega) + 
\tC_-(\omega) \delta \tC_-^*(\omega)   \right] \cr
&- \frac1{i\pi \k} \delta  \left[\int_0^{+\infty} \rd\omega \omega\left[  
\tC_-^*(\omega) \tC_-(\omega) \right]\right]\, ,
\ee
where $\widetilde{C}_{\pm}(\omega)$ are positive energy modes at $\mathscr{I}^+$,
\be 
C(u) = \frac{1}{2\pi}\int_0^{\infty} d\omega \left[e^{-i\omega u} \widetilde{C}_+(\omega) + e^{i\omega u} \widetilde{C}_-^*(\omega) \right].
\ee
Therefore we see that, after a canonical transformation generated by the number of negative helicity gravitons, namely $\alpha_- = \frac{i}{\pi \k} \int_0^{+\infty} \rd\omega \omega\left[  
\tC_-^*(\omega) \tC_-(\omega) \right] $, the symplectic potential can be 
written as $\theta=\theta_++\theta_- $ where
\be
\label{theta-omega}
\theta_\pm =  \frac1{i\pi\k} \int_0^{+\infty} \rd\omega \omega\left[  \tC_\pm(\omega) \delta \tC_\pm^*(\omega) \right]\,.
\ee
Note that shifts of the symplectic potential that are exact on phase space do not affect the symplectic form and, hence, the canonical commutation relations. 

We could equivalently use the fundamental reconstruction theorem to rewrite \eqref{Fourier-symplectic} in terms of the soft residues, namely 
\be
\theta_\pm &= \frac1{\pi\k} \int_0^{\infty} \rd\omega  \tN_\pm(\omega) \delta \tC_\pm^*(\omega) 
= \frac1{\pi\k}\int_0^{\infty} \rd\omega  \left(\sum_{n=0}^\infty \mathscr{M}_\pm(n) \omega^{n}\right) \delta \tC^*_{\pm}(\omega) \cr
&=
\frac1{\pi\k} \sum_{n=0}^\infty \mathscr{M}_\pm(n)   \left( \int_0^{\infty} \rd\omega  \omega^{n} \delta \tC_\pm^*(\omega) \right)\cr
&= {\f1{\pi i \k}}\sum_{n=0}^\infty \mathscr{M}_\pm(n) \delta \mathscr{S}_\pm^*(n),\label{thetapm}
\ee
where in the last line we used \eqref{Nh} and \eqref{Goldstones}.

The conjugates of the Goldstone operators are the conformally soft operators. We could have also seen this by using the Taylor expansion of $C(u)$ and  evaluating
\be \label{Directe}
\theta_+ =  \frac{1}{\pi i\kappa^2}\sum_{n=0}^\infty  \left(\frac{i^n}{n!} \int_{-\infty}^{+\infty} \rd u \,u^n N_+(u)\right)\delta \mathscr{S}^*_+(n)
= \frac{1}{\pi i \kappa^2}\sum_{n=0}^\infty  \mathscr{M}_+(n)\delta \mathscr{S}^*_+(n)\,,
\ee
in agreement with \eqref{thetapm}. In the last equality we used \eqref{C+} and \eqref{softC}.
Similarly, one finds
\be 
\theta_- = \frac{1}{\pi i \kappa^2}\sum_{n=0}^\infty  \mathscr{M}_-(n)\delta \mathscr{S}^*_-(n).
\ee
 Note that the relation \eqref{Nh} between $\widehat{N}$ and $\widehat{C}$ implies that
$\mathscr{M}(n)$ have dimensions $1-n$ while $\mathscr{S}(n)$ have dimensions $1+n$. 

\section{$R$-matrix}\la{App:R}

In this appendix we provide first the derivation of  \eqref{deltahat},  
\eqref{deltaalpha}. 
By plugging \eqref{Schnhat} into the definition \eqref{deltahatdef} and integrating by parts, we find
\be
\widehat\delta_{mn}(\alpha)& =
\frac{1}{\sqrt{ 2\pi\alpha }}
\f{i^{m-n}}{(m-1)!}
\int_{-\infty}^{+\infty} (\pa_u^{n-1} u^{m-1}) \,   e^{-\frac{u^2}{2\alpha}} \rd u\,,
\ee
which thus vanishes for $n>m$. Hence, assuming $n\leq m$, we arrive at 
\be
\widehat\delta_{mn}(\alpha)& =
\frac{1}{\sqrt{ 2\pi\alpha }}
\f{i^{m-n}}{(m-n)!}
\int_{-\infty}^{+\infty}  u^{m-n}    e^{-\frac{u^2}{2\alpha}} \rd u
\cr
&=\f{(1+(-)^{m-n})}2
i^{m-n}\left(2\alpha\right)^{\f{m-n}{2}}
\f{\Gamma\left(\f{m-n+1}{2}\right)}{\sqrt{ \pi }\Gamma(m-n+1)}
\cr
&= \pi_{m-n}\left(-\frac{\alpha}{2}\right)^{\frac{(m-n)}{2}} \f1{\Gamma\left(\frac{m-n}{2}+1\right)},
\ee
where we used the identity 
\be
\f{\Gamma\left(\f{m-n+1}{2}\right)}{\sqrt{ \pi }\Gamma(m-n+1)}
=\f{1}{2^{m-n}\Gamma\left(\frac{m-n}{2}+1\right)},
\ee
and defined
\be
\pi_{m-n}:=
\left \{
\begin{matrix}
0 &  n>m\\
\f{(1+(-)^{m-n})}2& n\leq m
\end{matrix}\right.
.
\ee

We now consider the Schwartzian state \eqref{Psiomega} and plug it into \eqref{dalpha} to derive \eqref{deltaalpha} as
\be
\delta_{nm}(\alpha)&=
\sqrt{ \frac{2}{\pi \alpha }}
\f{m! c_n}{n!(m-n)!}
\int_0^\infty \omega^{m-n}
\left(1-e^{-\frac{\omega^2}{2\alpha}}\right)^n
e^{-\frac{\omega^2}{2\alpha}} \rd \omega
\cr
&=
\sqrt{ \frac{2}{\pi \alpha }}
\f{m!c_n}{n!(m-n)!}
\sum_{k=0}^n
(-)^k\left (
\begin{matrix}
n \\
k
\end{matrix}
\right)
\int_0^\infty \omega^{m-n}
e^{-\frac{\omega^2(k+1)}{2\alpha}} \rd \omega
\cr
&=
\frac{1}{ \sqrt{ \pi  }}
\f{m!c_n}{n!(m-n)!}
 (2\alpha)^{\f{ m - n}{2}}
 \Gamma\left(\f{ m - n+1}{2}\right)
 \sum_{k=0}^n
(-)^k\left (
\begin{matrix}
n \\
k
\end{matrix}
\right)(k+1)^{\f{ n-m -1}{2}}
 \quad \forall~ m-n\geq0.\cr
\ee
In the case $m<n$, we have
\be
\delta_{nm}(\alpha)&=\sqrt{ \frac{2}{\pi \alpha }}
(-)^{n-m}
\, \f{m!c_n}{n!}\int_0^\infty \pa_\omega^{n-m} \left[\left(1-e^{-\frac{\omega^2}{2\alpha}}\right)^n
e^{-\frac{\omega^2}{2\alpha}}\right]
\rd \omega
\cr
&=
\left.
\sqrt{ \frac{2}{\pi \alpha }}
(-)^{n-m}
\, \f{m!c_n}{n!} \pa_\omega^{n-m-1} \left[\left(1-e^{-\frac{\omega^2}{2\alpha}}\right)^n
e^{-\frac{\omega^2}{2\alpha}}\right]
\right|^\infty_{0}=0\,.
\ee


Next, we evaluate the unregularized Goldstone two-point function using \eqref{C+0} as follows
\bea 
\langle 0|\mathscr{S}_{\pm}(n, z) \mathscr{S}_{\pm}^{\dagger}(m, w)|0\rangle
& = & (2\pi)^2 i^{n-m} \left. \pa_u^n \pa_{u'}^{m-1} 
\langle 0|C_\pm(u, z) N_{\pm}^{\dagger}(u', w)|0\rangle \right|_{u=u'=0}\cr
&=& 2\pi i^{n-m+1} \frac{\kappa^2}{2} \left.\pa_u^n \pa_{u'}^{m-1} \frac{-i}{u-u'-i\epsilon}
\right|_{u=u'=0} \delta^2(z,w)\cr
&=& \kappa^2 \pi i^{n-m+1} (-1)^n (n+m-1)! \left.\frac{-i}{(u-u'-i\epsilon)^{n+m}}
\right|_{u=u'=0} \delta^2(z,w)\cr
&=& \kappa^2 \pi \frac{ \Gamma(n+m)}{\epsilon^{n+m} } \delta^2(z,w).
\eea 

On the other hand, upon regularization, we find that the Goldstone two-point function is proportional to 
\begin{equation}
\begin{split} \label{Rev}
R_{nm}(\alpha) &= \frac{i^{ m - n+1}}{\alpha}\int_{-\infty}^{\infty}du \p_u^{n - 1}e^{-\frac{u^2}{2\alpha}} \p_u^{m} e^{-\frac{u^2}{2\alpha}}\\
&= \int_{0}^{\infty} d\omega  \omega^{n + m - 1} e^{-\alpha \omega^2} = \frac{\Gamma(\frac{m+n}{2})}{\alpha^{\frac{m+n}{2}}},
 \end{split}
\end{equation}
where in the second line we have used the Fourier representation of the Gaussian projected onto the positive/negative energy components
\be
   \left.e^{-\frac{u^2}{2\alpha}}\right|_{\pm} = \frac{\sqrt{2\pi \alpha}}{2\pi}\int_0^{\infty}d\omega e^{-i\omega u} e^{-\frac{\alpha\omega^2}{2}}.
\ee

Finally, the regulated memory two-point function is proportional to
\be
    \widetilde{R}_{nm}(\alpha) = \frac{1}{(2\pi)^2}\int_0^{\infty} d\omega \omega  \Psi_{n,\alpha}^*(\omega) \Psi_{m,\alpha}(\omega),
\ee
where $\Psi_{n,\alpha}(\omega)$ is given in \eqref{Psiomega}. This can be written in terms of the Fourier transform
\be
\begin{split}
\widetilde{\Psi}_{n,\alpha}(u) &= \frac{1}{2\pi}\int_0^{\infty}d\omega e^{-i\omega u} \Psi_{n,\alpha}(\omega) = \frac{1}{2\pi} 2\sqrt{\frac{2\pi}{\alpha}}\frac{1}{n!}\int_0^{\infty}d\omega (-iu)^n e^{-i\omega u} c_n(1 - e^{-\frac{\omega^2}{2\alpha}})^n e^{-\frac{\omega^2}{2\alpha}}\\
&= \sum_{k = 0}^n \left(\begin{matrix}n\\
k
\end{matrix}\right) \frac{(-1)^k}{n!\sqrt{1 + k}} e^{-\frac{\alpha u^2}{2(1 + k)}}(-iu)^n c_n\left(1 - {\rm  Erf}\left[i\frac{\sqrt{\alpha} u}{\sqrt{2(1+k)}}\right] \right)
\end{split}
\ee
as
\be
\widetilde{R}_{nm}(\alpha) = \frac{1}{2\pi}\int_{-\infty}^{\infty} du \widetilde{\Psi}_{n,\alpha}^*(u) (i\p_u) \widetilde{\Psi}_{m,\alpha}(u).
\ee
We can now change variables $x = \sqrt{\alpha} u$ in which case the $\alpha$ dependence factorizes and we are left with
\be
\begin{split}
\widetilde{R}_{nm}(\alpha) &= \frac{C_{nm}}{\alpha^{\frac{m + n}{2}}},
\end{split}
\ee 
where the $\alpha$-independent coefficient is
\be
\label{Cnm}
\begin{split}
C_{nm} &= \frac{1}{2\pi}\sum_{k=0}^n \sum_{k' = 0}^m  \left(\begin{matrix}n\\
k
\end{matrix}\right) \left(\begin{matrix}m\\
k'
\end{matrix}\right) \frac{i^{n-m +1}(-1)^{k + k'}}{n!m!\sqrt{(1 + k)(1 + k')}}\\
&\times \int_{-\infty}^{\infty} dx e^{-\frac{x^2}{2(1+k)}} c_n^* c_m x^n \left(1 - {\rm Erf}\left[\frac{ix}{\sqrt{2(1 + k)}}\right]\right) \p_x\left[x^m e^{-\frac{x^2}{2(1+k')}}\left(1 - {\rm Erf}\left[\frac{ix}{\sqrt{2(1 + k')}}\right]\right)\right].
\end{split}
\ee

\section{Delta function identity}
\label{sec:completeness}

Consider a Schwartzian function $f(u) \in \mathcal{S}$ and  define
\be 
\begin{split}
\mathscr{F}_{+}(u,v) &:= \frac{1}{2i\pi} \sum_{n \geq 0} \left(\frac{ v+i \epsilon }{u - i\epsilon} \right)^{n} \frac{1}{u - i\epsilon} .
\end{split}
\ee
We also denote $\mathscr{F}(u,v)=\mathscr{F}_+(u,v)+\mathscr{F}_+(v,u)$.
In this section we show that when $f\in \mathcal{S}$ we have 
\be \la{Ff}
\int_{-\infty}^{\infty} \rd v\, \mathscr{F}_+(u,v) f(v) =  f_+(u), \qquad
\int_{-\infty}^{\infty} \rd v\, \mathscr{F}_{{+}}({v,u}) f(v) =  f_-^*(u),
\ee
where we denote by $f_\pm(u)$ the positive and negative energy component of $f(u)=f_+(u)+f_-^*(u)$.
These are defined according to  the formulae (\ref{C+},\ref{C-}).

In order to prove the first equality in \eqref{Ff} we use \eqref{C+} to get an asymptotic expansion of $f_+(u)$ around $u=\infty$\footnote{This expansion is ambiguous as the $i\epsilon$ could be split  differently as 
\bea
\frac1{2i\pi} \sum_{n=0}^\infty \frac1{(u-i\epsilon)^{n+1}} \left[\int_{-\infty}^{+\infty} \rd v v^{n} f(v)  \right]\,~~\mathrm{or}\, 
~~\frac1{2i\pi} \sum_{n=0}^\infty \frac1{u^{n+1}} \left[\int_{-\infty}^{+\infty} \rd v (v+i\epsilon)^{n} f(v)  \right].
\eea
In the main text we choose the $i\epsilon$ prescription in the denominators since for all practical purposes $(v+i\epsilon)^n$ is equal to $v^n$.
}
\be
f_+(u) &=  \frac1{2i\pi}  \int_{-\infty}^{+\infty} \rd v \frac{f(v)}{[(u-i\epsilon)- v]} = 
\frac1{2i\pi} \sum_{n=0}^\infty \frac1{(u-i\epsilon)^{n+1}} \left[\int_{-\infty}^{+\infty} \rd v v^{n} f(v)  \right].
\ee 
The existence, the integral coefficients and the convergence of the sum for large $u$ follows from the assumption that $f$ is Schwartzian.

 Similarly, to prove the second equality in \eqref{Ff}, we use  \eqref{C-} to  get a Taylor expansion of $f_-(u)$ near $u=0$. First one establishes that 
 \be
\pa_u^nf_-^*(u) &=  \frac{n!}{2i\pi}  \int_{-\infty}^{+\infty} \rd v \frac{f(v)}{(v-u-i\epsilon)^{n+1}},\la{C++}
\ee
which implies  
\be 
f_-^*(u) &= \frac{1}{2i\pi} \sum_{n=0}^\infty u^n \left[ \int_{-\infty}^{+\infty} \rd v \frac{f(v)}{(v-i\epsilon)^{n+1}}\right].\la{C--}
\ee 
Given these results, 
we can now straightforwardly conclude that 
\be 
\begin{split}
\int_{-\infty}^{\infty} \rd v\, \mathscr{F}(u,v) f(v) &= \frac1{2i\pi}\sum_{n \geq 0} \frac{i}{(u -i \epsilon)^{n+1}} \int_{-\infty}^{\infty} \rd v\, v^n f(v) + 
\frac1{2i\pi} \sum_{n \geq 0}  \left[ \int_{-\infty}^{+\infty} \rd v \frac{u^n f(v)}{(v-i\epsilon)^{n+1}}\right] \\
&= f_+(u) + f_-^*(u) = f(u).
\end{split}
\ee

\section{All order dressing identity}
\label{app:dress}
\label{app:dress}
In this section we want to evaluate
\be
I_{ij}^{nm} :=
\int \rd^2w \frac{( z_i-w)^n}{( \bz_i-\bw)}  \frac{( \bz_j-\bw)^m}{( z_j-w)} = (2\pi)^2 n! m! \int \rd^2w G^+_{n+1}(w,z_i)
G^-_{m+1}(w,z_j).
\ee
The second expression shows that this integral satisfies the identities
\be \label{Greens}
\pa^{n+1}_{z_i} I_{ij}^{nm} = 2\pi  n ! \frac{(\bz_i-\bz_j)^m}{(z_i - z_j)}, \qquad
\pa^{m+1}_{\bz_j} I_{ij}^{nm} = 2\pi   m ! \frac{(z_j-z_i)^n}{(\bz_j - \bz_i)}.
\ee 

We see that we can write \eqref{Iij} as
\be 
I_{ij}^{nm} = \frac{1}{n!} \int \rd^2z (z_i - z)^n \p_{\bz}^{n + 1}\left((\bz - \bz_i)^n \log|z - z_i|^2 \right) \frac{(\bz_j - \bz)^m}{(z - z_j)}.
\ee
The validity of this expression follows from the recursion relation
\be  
\label{rec-rel}
\pa^{n+1}(z^n \log|z|^2)= 
\pa^{n}(nz^{n-1} \log|z|^2 + z^{n-1})
= n \pa^{n}(z^{n-1} \log|z|^2 )= \frac{n!}{z},
\ee 
or alternatively, by direct computation
\be 
\p_{\bz}^{n+1}\left[(\bz - \bz_i)^n\log|z - z_i|^2 \right] = \sum_{p = 1}^{n+1} \frac{(-1)^{p-1}(n+1)!n!}{p!(n - p+1)!} \frac{1}{\bz - \bz_i} = \frac{n!}{\bz - \bz_i}.
\ee
Now, integrating by parts and assuming wlog that $n \geq m$,  we find 
\be 
\label{Imn}
I_{ij}^{n m}(z_i,z_j) = -2\pi (z_i - z_j)^{n}(\bz_{j} - \bz_{i})^{m} \log\left( |z_i - z_j|^2 H_{nm} \right).
\ee
 $H_{nn}=1$ while in general $H_{nm}$ is given by  
\be\label{defA}
\log H_{nm}= |H_n-H_m|,
\ee
where $H_n:=\sum_{k=1}^{n}\frac1k$ are the harmonic numbers. One can directly check that \eqref{Imn} satisfies \eqref{Greens}. 

This is obtained as follows (for $n \geq m$)
\be
\begin{split}
   I^{nm}_{ij} &= \frac{1}{n!}\int d^2z (z_i - z)^n \p_{\bz}^{n+1}\left((\bz - \bz_i)^n \log|z - z_i|^2 \right) \frac{(\bz_j - \bz)^m}{z - z_j}\\
   &= -\frac{m!}{n!}\int d^2z (z_i - z)^n \p_{\bz}^{n-m}\left((\bz - \bz_i)^n \log|z - z_i|^2 \right) 2\pi \delta^{(2)}(z - z_j)\\
   &= -2\pi (z_i - z_j)^{n}(\bz_{j} - \bz_i)^{m} \log |z_i - z_j|^2 -  2\pi \left(\sum_{k = m+1}^{n}  \frac{1}{k}\right) (z_i - z_j)^n(\bz_j - \bz_i)^m,
   \end{split}
\ee
where in the last line we used the generalization of \eqref{rec-rel} to arbitrary powers of derivatives
\be
\begin{split}
\p^n\left(\frac{z^m}{m!} \log|z|^2\right) &= \p^{n-1}\left( \frac{z^{m-1}}{(m-1)!}\log|z|^2 + \frac{z^{m-1}}{m!} \right) = \p^{n-1}\left( \frac{z^{m-1}}{(m-1)!}\log|z|^2 \right) + \frac1{m} \frac{z^{m - n}}{(m - n)!}\cr 
    &= \left( \frac{z^{m-n}}{(m-n)!}\log|z|^2 \right) + \left(\sum_{k=0}^{n-1}\frac1{m-k} \right) \frac{z^{m - n}}{(m - n)!}.
\end{split}
\ee
The polynomial contribution is absent for $n = m$. 

\subsection{Unitarity of dressing}
\label{sec:dress-unitary}
In this section we study the hermiticity property of the  soft operators \eqref{eq:msout}.
We prove that 
\be 
\left[m_+^{(s)}(z;w)\right]^\dagger = (-1)^s m_-^{(s)}(z;w).
\ee 
The Klein-Gordon inner product \eqref{KGprod} is given by 
\be 
\langle f| g \rangle &\equiv 
-i\frac{2}{\kappa^2} \int du \int_{S^2} d^2 w \pa_u f^*(u,w) g(u,w) = i \frac{2}{\kappa^2}\int du \int_{S^2} d^2 w  f^*(u,w) \pa_u g(u,w).
\ee  
We consider 
\be 
\langle f| m_+^{(s)}(z;w)| g \rangle \equiv \langle f| m_+^{(s)}(z;w)[g] \rangle = -i\frac{2}{\kappa^2}\int du \int_{S^2} d^2 w \pa_u f^*(u,w) m_+^{(s)}(z;w) [g(u,w)].
\ee
And we evaluate 
\be 
&  \quad i^{s} \langle f| m_+^{(s)}(z;w) | g \rangle
\\
&=\frac{1}{2} 
  \sum_{\ell = 0}^s
  \int du \int_{S^2} d^2 w \p_{w}^{s - \ell} \delta^{(2)}(z, w) \pa_u f^*(u,w)
  \frac{( \ell +1)(\widehat\Delta + 2)_{s - \ell}}{(s - \ell )!}    \p_{w}^\ell
  \pa_u^{1-s }g(u,w)
  \cr
&=\frac{1}{2} 
 \sum_{\ell = 0}^s (-1)^{ \ell }
  \int du \int_{S^2} d^2 w
    \p_{w}^\ell ( \p_{w}^{s - \ell} \delta^{(2)}(z, w) \pa_uf^*(u,w) )
  \frac{( \ell +1)(\widehat\Delta + 2)_{s - \ell}}{(s - \ell )!}  
   \pa_u^{1-s }g(u,w) 
  \cr
&=\frac{1}{2} 
 \sum_{\ell = 0}^s\sum_{p=0}^\ell (-1)^{ \ell }
  \int du \int_{S^2} d^2 w
    \p_{w}^p \pa_uf^*(u,w) \p_{w}^{s - p} \delta^{(2)}(z, w) 
  \frac{( \ell +1)(\widehat\Delta + 2)_{s - \ell}}{(s - \ell )!}  
  \f{(\ell)_p}{p!}
   \pa_u^{1-s }g(u,w)  
  \cr
&=\frac{1}{2} 
\sum_{p=0}^s
  \int du \int_{S^2} d^2 w
    \p_{w}^{s - p} \delta^{(2)}(z, w)
   \p_{w}^p\pa_u f^*(u,w) 
\left[ \sum_{\ell = p}^s   (-1)^{ \ell } \frac{( \ell +1)(\widehat\Delta + 2)_{s - \ell}}{(s - \ell )!}   \f{(\ell)_p}{p!} \right]
   \pa_u^{1-s }g(u,w) \nonumber   \,.
\ee
We can simplify this expression using the binomial identity
\be 
\frac{(x-\ell-1)_s}{s!}=\sum_{n=0}^s (-)^n
 \frac{(\ell+n)_{n}}{n!} \frac{(x)_{s-n}}{(s-n)!},
\ee
which allows us to evaluate
\be
\sum_{\ell = p}^s   (-1)^{ \ell } \f{( \ell +1) (\ell)_p}{p!} \frac{(\widehat\Delta + 2)_{s - \ell}}{(s - \ell )!}    &= (-1)^p(p+1)\sum_{n = 0}^{s-p}   (-1)^{ n }  \f{(n+p+1)_{p+1}}{(p+1)!} \frac{(\widehat\Delta + 2)_{s - \ell}}{(s - \ell )!}  
\cr
&=(-1)^{p} (1 + p)
\f{(  \widehat\Delta  -p)_{s-p}}{(s-p)!}\,.
\ee
To continue the evaluation we take advantage of the  useful identities
\be
&u^n\p_u^n=(\widehat\Delta-1)_n\,,\qquad \p_u^nu^n=(\widehat\Delta+n-1)_n\,,\qquad
u^{-n}\p_u^{-n}=(\widehat\Delta+n-1)^{-1}_n\,,\cr
& \p_u (\widehat\Delta+\alpha)_n= (\widehat\Delta+\alpha+1)_n \p_u\,,\qquad
\p_u^{-1} (\widehat\Delta+\alpha)_n= (\widehat\Delta+\alpha-1)_n \p_u^{-1}\,,
\label{uDelta}
\ee
which allow us to evaluate
\be
\left[(  \widehat\Delta  -p)_{s-p}\right]^\dagger
&=
\left[\pa_u^{1-p}(  \widehat\Delta-1 )_{s-p} \pa_u^{p-1}\right]^\dagger
 \cr
&=  \pa_u^{p-1} \left[u^{s-p}\pa_u^{s-p} \right]^\dagger \pa_u^{1-p} 
\cr
&= (-1)^{s-p}
 \pa_u^{p-1} \left[\pa_u^{s-p} u^{s-p}\right] \pa_u^{1-p}\cr
 &=(-1)^{s-p}
 \pa_u^{p-1} (  \widehat\Delta +s -p-1 )_{s-p} \pa_u^{1-p}\cr
& =(-1)^{s-p}  
 (  \widehat\Delta +s -2 )_{s-p}.
\ee 
We can thus write
\be 
& i^{s} \langle f| m_+^{(s)}(z;w) | g \rangle \cr
& =\frac{1}{2} 
 \sum_{p=0}^s (-1)^{ p }
  \int du \int_{S^2} d^2 w
   \p_{w}^{s - p} \delta^{(2)}(z, w)  \p_{w}^p\pa_uf^*(u,w) 
   \left[
\f{ (1 + p) (  \widehat\Delta  -p)_{s-p}}{(s-p)!}
\right]\pa_u^{1-s}
  g(u,w) 
  \cr
  &= 
  \frac{1}{2} 
 \sum_{p=0}^s (-1)^{ p-s }
  \int du \int_{S^2} d^2 w
   \p_{w}^{s - p} \delta^{(2)}(z, w)
  \pa_u g(u,w)     \pa_u^{-s}
 \left[
\f{ (1 + p) (  \widehat\Delta  -p)_{s-p}}{(s-p)!}
\right]^\dagger  \p_{w}^p \pa_u f^*(u,w) 
  \cr
 &=  
  \frac{1}{2} 
 \sum_{p=0}^s
  \int du \int_{S^2} d^2 w
   \p_{w}^{s - p} \delta^{(2)}(z, w)
  \pa_u g(u,w)     \pa_u^{-s}
 \left[
\f{ (1 + p) (  \widehat\Delta +s-2)_{s-p}}{(s-p)!}
\right]  \p_{w}^p \pa_u f^*(u,w) 
  \cr
  &=  
  \frac{1}{2} 
 \sum_{p=0}^s
  \int du \int_{S^2} d^2 w
   \p_{w}^{s - p} \delta^{(2)}(z, w)
  \pa_u g(u,w)     
 \left[
\f{ (1 + p) (  \widehat\Delta -2 )_{s-p}}{(s-p)!}
\right]  \p_{w}^p \pa_u^{1-s} f^*(u,w) 
  \cr
  &= \left[ 
  i^{s} \langle g | m_-^{(s)}(z;w) | f \rangle\right]^*.
\ee

\subsection{State factorization} 
\label{sec:state-factorization}
It will be convenient to decompose the operator into a product 
\be
\mathcal{D}(p) ={\cal N}(p) \mathcal{D}_{+}^\dagger(p) \mathcal{D}_-^\dagger(p) \mathcal{D}_{+}(p) \mathcal{D}_-(p),
\ee 
where ${\cal N}(p)$ is a normalization factor and 
where 
\be 
{\langle p|}\mathcal{D}_{\pm}(p) := \langle p|\exp\left\{\sum_{s =0}^{\infty} \frac{i^{- s}}{\pi \kappa^2} \int \rd^2 w \int \rd^2z  \left[ m^{(s)}_{\mp}( w;p)G_{s+2}^{\pm}(w;z)\mathscr{S}_{\pm}(s,z) \right] \right\}.
\ee
These operators are such that
\be
\quad [\mathcal{D}_+(p_i), \mathcal{D}_-(p_j)] = 0, \quad  i \neq j.
\ee
To compute correlation functions for a product of exponential operators, we need to use the Baker--Campbell--Hausdorff formula 
\be 
\label{eq:BCH}
e^{A + B} = e^A e^{B} e^{-\frac{1}{2}[A, B]},
\ee
valid when $[A, B]$ is a $c$-number, that allows us to derive an important identity obeyed by coherent states of the form
\be 
\label{eq:toy-dress}
\mathcal{W}_i = \exp\left\{\alpha^i_-(a_+ - a_-^{\dagger}) + \alpha_+^i(a_- - a_+^{\dagger}) \right\}.
\ee
Here $a_{\pm}, a^{\dagger}_{\pm}$ are harmonic oscillators while $\alpha^i_{\pm}$ are (possibly operator-valued) coefficients that commute with the oscillators, as well as each other for $i \neq j$. Note that such coherent states mimick our dressings \eqref{eq:Dress} and, using \eqref{eq:BCH}, can be equivalently written as
\be 
\mathcal{W}_i = e^{-\alpha^i_- a_-^{\dagger}} e^{-\alpha^i_+ a_+^{\dagger}} e^{\alpha^i_- a_+} e^{\alpha^i_+ a_-} e^{-\frac{1}{2} \alpha_+^i \alpha_-^i([a_+, a_+^{\dagger}] + [a_-, a_-^{\dagger}])}.
\ee

Correlation functions of coherent states \eqref{eq:toy-dress} are then evaluated by successively moving all of the $a_{\pm}^{\dagger}$ to the left, namely
\be 
\label{eq:toy-correlator}
\langle 0|\mathcal{W}_1 \cdots \mathcal{W}_n |0\rangle = \exp\left\{-\sum_{i < j}\alpha_+^i \alpha_-^j\left([a_+, a_+^{\dagger}] + [a_-, a_-^{\dagger}] \right) \right\} \exp\left\{ - \sum_{i = 1}^n \frac{1}{2} \alpha_-^i\alpha_+^i( [ a_+,  a_+^{\dagger}] +  [ a_-, a_-^{\dagger}]) \right\}.
\ee

\section{Charges} \la{App:Q}
In this appendix we collect technical aspects concerning the formulation of the tower of higher spin charges in terms of the discrete tower of Goldstone and memory operators. 

\subsection{Quadratic charges}\la{App:Q2}

The quadratic higher spin charge aspects are obtained from the general relation \eqref{qren}
as
\be\la{Q2ren}
\hat q^2_s(u,z)
&=\frac14\p_u^{-1}  \left[\sum_{\ell=0}^s \frac{(\ell+1) (-u)^{s-\ell}}{(s-\ell)!} 
  D^{s-\ell}\left[C  D^{\ell} \p_u^{-\ell+1}  N  \right](u,z) \right]\,.
\ee

We can further simplify this expression by defining the operator $\widehat\Delta :=\pa_u u$ and using that $u^n\p_u^n=(\widehat\Delta-1)_n$, where 
$ (x)_n= x(x-1)\cdots(x-n+1)$ is the falling factorial.
This allows us to rewrite
\be
 u^{s-\ell}  \p_u^{-\ell+1}  N  =  (\hD -1)_{s-\ell} \pa_u^{1-s} N,
\ee
and therefore we get the charge expression
\be 
Q_s^2(\tau) = {-}\frac14 
\sum_{\ell=0}^s \frac{(\ell+1) }{(s-\ell)!}  \int_{\scri} 
  \left[ C \left(D^{\ell} (\hD -1)_{s-\ell} \p_u^{1-s}  N \right)  \right](u,z)  D^{s-\ell} \tau_s(z).
\ee
This gives us the first line of \eqref{Qs}. 
Integrating by parts and using that 
\be
 \int_{-\infty}^{+\infty} \rd u \hD A(u) B (u)= \int_{-\infty}^{+\infty} \rd u  A(u) (1-\hD) B(u)\,,
 \ee
  we get the alternate expression
\be
Q_s^2(\tau) &={-} \frac14  \sum_{\ell=0}^s   \frac{(\ell+1) }{(s-\ell)!} 
\int_{\scri}  
   \left(D^{\ell} (\hD -1)_{s-\ell} \p_u^{1-s}  N(u,z)  \right)   \left[ C(u,z)   D^{s-\ell} \tau_s(z)\right] \cr
&= {-}\frac14 \sum_{\ell=0}^s \frac{(\ell+1) }{(s-\ell)!}(-1)^{\ell-s} \int_{\scri} 
    {C^*}(u,z)     D^{\ell} \left[  \left( \p_u^{2-s} (-\hD )_{s-\ell} C(u,z)  \right)  D^{s-\ell} \tau_s(z) \right], \cr
&= {-}\frac14 \sum_{\ell=0}^s \frac{(\ell+1) }{(s-\ell)!}
    \int_{\scri} 
    {C^*}(u,z)     D^{\ell} \left[  \left( (\hD +1-\ell)_{s-\ell} \p_u^{1-s}{N^*}(u,z)  \right)  D^{s-\ell} \tau_s(z)\right] \cr
 &={-} \frac14 \sum_{p=0}^{s}
\left[\sum_{\ell=p}^s  \frac{ (\ell+1)!}{p! (\ell-p)!(s-\ell)!}
 \right] 
 \int_{\scri}    {C^*}(u,z)       D^{p}\left( (\hD +1-\ell)_{s-\ell} \p_u^{1-s}{N^*}(u,z)  \right)  
    D^{s-p} \tau_s(z)\,,\cr
    \ee
    where we used $\pa_u^{-1} \hD \pa_u= \hD-1$  and  
    $(-x)_n = (-1)^n ( x+n-1)_n$ in the third equality,   which implies that 
    $\p_u^{2-s} (-\hD )_{s-\ell}  = (-1)^{s-\ell} (\hD+1-\ell)_n\pa_u^{2-s}$;
     we also expanded the derivative $D^\ell$ in the fourth equality.
    To evaluate this expression one uses the identity 
    \be
    \sum_{\ell=p}^s  \frac{ (\ell+1)!}{ (\ell-p)!(s-\ell)!} (\widehat{\Delta}-\ell)_{s-\ell} = \frac{(p+1)!}{(s-p)!} (\widehat{\Delta}+2)_{s-p},
    \ee
    which means that 
    \be
    Q_s^2(\tau) = - \frac14 \sum_{p=0}^{s} \frac{(p+1)}{(s-p)!}
    \int_{\scri}  \left[  {C^*}  D^{p}\left( (\hD +3)_{s-p} \p_u^{1-s}{N^*} \right)  \right](u,z)
    D^{s-p} \tau_s(z) \,,\la{Q2conj}
    \ee
 and we obtain the second line of \eqref{Qs}.

To re-express the quadratic term \eqref{Q2ren} in terms of the corner variables $ \mathscr{M}_\pm(n,z), \mathscr{S}_\pm(n,z)$, we start with
\be 
Q_s^2(\tau) &= -\frac14 
\sum_{\ell=0}^s \frac{(\ell+1) }{(s-\ell)!}  \int_{\scri}   D^{s-\ell} \tau_s(z)
  \left[ (C_+(u,z)+ C_-^*(u,z))  \left(D^{\ell} (\hD -1)_{s-\ell} \p_u^{1-s}   ({N}_-(u,z)+ {N}_+^*(u,z))  \right)  \right] 
  \cr
 &=- \frac14 
\sum_{\ell=0}^s \frac{(\ell+1) }{(s-\ell)!}  \int_{\scri}   D^{s-\ell} \tau_s(z)
  \bigg[ C_+(u,z) D^{\ell} (\hD -1)_{s-\ell} \p_u^{1-s}  {N}_+^*(u,z) 
  \cr
  &+ C_-^*(u,z)  D^{\ell} (\hD -1)_{s-\ell} \p_u^{1-s}   {N}_-(u,z)  \bigg].\nonumber 
\ee 
We thus see that again we can split the quadratic charge aspects as \eqref{Qchargesplit} into
\be\la{Q2split}
 Q^2_s(\tau) =
  -\frac1{4}\left[ i^s Q^2_-(s)  + i^{-s}Q_+^{*2}(s)\right],
\ee
where
\be
Q^{2*}_{s+ }(\tau)&=
 \frac {i^{1+s}}{2\pi} 
\sum_{n=0}^\infty \sum_{\ell=0}^s
\frac{i^{- n}}{n!} 
\frac{(\ell+1) }{(s-\ell)!}  \int_{\scri}   D^{s-\ell} \tau_s(z)
\mathscr{S}_+(n,z) u^n D^{\ell} (\hD -1)_{s-\ell} \p_u^{1-s}  {N}_+^*(u,z) 
\cr
&=
 \frac {i^{1+s}}{2\pi} 
\sum_{n=0}^\infty \sum_{\ell=0}^s
\frac{i^{- n}}{n!} 
\frac{(\ell+1) }{(s-\ell)!}  \int_{S}  D^{s-\ell} \tau_s(z)
\mathscr{S}_+(n,z)   ( -n-1)_{s-\ell}D^{\ell} \int_{-\infty}^\infty \rd u u^n \p_u^{1-s}  {N}_+^*(u,z) 
\cr
&=
- \frac {i^{1+s}}{2\pi}  
\sum_{n=0}^\infty \sum_{\ell=0}^s
\frac{(-)^{s+\ell+1}i^{- n}}{n!} 
\frac{(\ell+1) (n+s-\ell)_{s-\ell}}{(s-\ell)!} 
\cr
&\times
\int_{S}   D^{s-\ell} \tau_s(z)
\mathscr{S}_+(n,z)   D^{\ell} \int_{-\infty}^\infty \rd u  u^{n+s-1} (\hD +s-2)^{-1}_{s-1}  {N}_+^*(u,z) 
\cr
&=
- \frac {i^{1+s}}{2\pi} 
\sum_{n=0}^\infty \sum_{\ell=0}^s
\frac{(-)^{s+\ell+1}i^{- n}(\ell+1)(n+s-\ell)_{s-\ell} }{n! (-n-1)_{s-1} (s-\ell)!} 
\cr
&\times \int_{S}   D^{s-\ell} \tau_s(z)
\mathscr{S}_+(n,z)   D^{\ell} \int_{-\infty}^\infty \rd u   u^{n+s-1}   {N}_+^*(u,z)
\cr
&=
 -\frac{1}{4\pi} 
\sum_{n=0}^\infty \sum_{\ell=0}^s
(-)^{\ell+s}(\ell+1)
 \left(\begin{matrix}
s+n-\ell\\
n
\end{matrix}\right)
\int_{S}    D^{s-\ell} \tau_s(z)
\mathscr{S}_+(n,z)   D^{\ell} \mathscr{M}^*_+(n+s-1,z).
\cr
\ee
 We have used \eqref{softC}, \eqref{uexp}
 and the identities
\be\la{uDelta}
%
\f1{ (-n-1)_{\ell}}=(-)^{\ell} \f{n!}{(n+\ell)!}\,,\qquad
 u(\hD+\alpha-1)^{\pm 1}_n=(\hD+\alpha-2)^{\pm 1}_n u\,,
\ee
valid $\forall~ n\geq 0\,,\alpha\in \Z$, and the fact  that the operator $\hD$  integrates to zero due to  choice of boundary conditions.

Let us perform an extra integration by parts to write
\be
Q^{2*}_{s+}(\tau)&=-
\frac{  (-)^{s}}{4\pi }  \sum_{n=0}^\infty \sum_{m=0}^s \sum_{\ell=m}^s  (\ell+1)
\f{(n+s-\ell)_{s-\ell}}{(s-\ell)!}
\f{(\ell)_m}{m!}
 \int_S  D^{s-m}\tau_s(z)   {\mathscr{M}}^*_+(s+n-1,z) D^{m}\mathscr{S}_+(n,z)
 \cr
 &=-
\frac{  (-)^{s}}{4\pi }   \sum_{n=0}^\infty \sum_{m=0}^s   (m+1)
\f{(n+s+2)_{s-m}}{(s-m)!}
 \int_S   D^{s-m}\tau_s(z)   {\mathscr{M}}^*_+(s+n-1,z) D^{m}\mathscr{S}_+(n,z)\,,
\ee
where we used
\be
\sum_{\ell=m}^s  (\ell+1)
\f{(n+s-\ell)_{s-\ell}}{(s-\ell)!}
\f{(\ell)_m}{m!}
&=
 (m+1)
 \f{(n+s-m)_n}{n!}
{}_2F_1[2 + m, m - s, m - n - s, -1]
\cr
&=  (m+1)
\f{(  n + s+2 )_{s-m}}{(s-m)!}\,.
\ee
Analogous expressions are obtained for $  Q^2_{s-}(\tau)$  in terms of $\mathscr{S}_-(n), \mathscr{M}^*_-(s+n-1) $.

It is also useful to derive the corner variables expression of the quadratic charge starting from the version \eqref{Q2conj} in terms of the complex conjugate fields. We have
 \be
    Q_s^2(\tau) &=-  \frac14 \sum_{p=0}^{s} \frac{(p+1)}{(s-p)!}
    \int_{\scri}   D^{s-p} \tau_s(z)  (C^*_+(u,z)+ C_-(u,z))  D^{p}\left( (\hD +3)_{s-p} \p_u^{1-s}({N}^*_-(u,z)+ {N}_+(u,z)) \right)  
    \cr
    &= -\frac14 \sum_{p=0}^{s} \frac{(p+1)}{(s-p)!}
    \int_{\scri}  D^{s-p} \tau_s(z) \bigg[
    C^*_+(u,z)  D^{p}\left( (\hD +3)_{s-p} \p_u^{1-s} {N}_+(u,z) \right) \cr
&    +C_-(u,z))  D^{p}\left( (\hD +3)_{s-p} \p_u^{1-s}({N}^*_-(u,z)
        \right) 
    \bigg]\,.\nonumber
    \ee
  Hence, using again the split \eqref{Q2split},
  we have for the positive component
\be
Q_{s+}^{2*}(\tau)&=
-\frac {i^{s+1}}{2\pi} \sum_{\ell=0}^{s}\sum_{n=0}^\infty \frac{(\ell+1)}{(s-\ell)!}
 \frac{(-i)^{-n}}{n!}
    \int_{\scri}  D^{s-\ell} \tau_s(z)
     \mathscr{S}^*_+(n)  D^{\ell}\left( u^n (\hD +3)_{s-\ell} \p_u^{1-s} {N}_+(u,z) \right)
     \cr
&=\frac {i^s}{2\pi} \sum_{\ell=0}^{s}\sum_{n=0}^\infty \frac{(\ell+1)}{(s-\ell)!}
 \frac{i^{n-1}} {n!}
    \int_{\scri}  D^{s-\ell} \tau_s(z)
     \mathscr{S}^*_+(n)  D^{\ell}\left(  (3-n)_{s-\ell} u^{n+s-1} (\hD +s-2)^{-1}_{s-1} {N}_+(u,z) \right)   
     \cr
&=\frac {i^s}{2\pi}\sum_{\ell=0}^{s}\sum_{n=0}^\infty \frac{(\ell+1)}{(s-\ell)!}
 \frac{i^{n-1}}{n!}
 \f{(3-n)_{s-\ell}}{  (-n -1)_{s-1}}
    \int_S  D^{s-\ell} \tau_s(z)
     \mathscr{S}^*_+(n)  D^{\ell} \int_{-\infty}^\infty \rd u u^{n+s-1} {N}_+(u,z)    
     \cr
&=-\frac{(-)^{s}}{4\pi} \sum_{\ell=0}^{s}\sum_{n=0}^\infty (\ell+1) \frac{ (3-n)_{s-\ell} }{(s-\ell)!}
    \int_S  D^{s-\ell} \tau_s(z)
     \mathscr{S}^*_+(n)  D^{\ell} \mathscr{M}_+(n+s-1)   \,.
\ee

Finally we compute the quantum action\footnote{We restore here explicit measure on the 2-sphere to avoid confusion between different sets of coordinates.}
\be
\begin{split}
[ Q^{2\dagger}_{s+}(\tau), \mathscr{S}_+(n,z) ]&=
-\frac{(-)^{s} }{4\pi}  \sum_{m=0}^\infty \sum_{\ell=0}^s  (-)^\ell(\ell+1)
   \left(\begin{matrix}
s+m-\ell\\
m
\end{matrix}\right)\cr
&\times
\int_{S}\rd^2z' \sqrt{q}\, D^{s-\ell}_{z'} \tau_s(z')
\mathscr{S}_+(m,z') D_{z'}^{\ell}  [{\mathscr{M}}^\dagger_+(s+m-1,z')  , \mathscr{S}_+(n,z) ]  \cr
 &=\f{\kappa^2}{4}(-)^{s}    \sum_{\ell=0}^s (-)^\ell (\ell+1)
   \left(\begin{matrix}
n+1-\ell\\
n+1-s
\end{matrix}\right)
\int_{S}\rd^2z' \sqrt{q}\, D^{s-\ell}_{z'} \tau_s(z')
  \mathscr{S}_+(n-s+1,z') D_{z'}^{\ell} \d^{(2)}(z,z')  \cr
 &=\f{\kappa^2}{ 4}(-)^{s}   \sum_{\ell=0}^s  \sum_{k=0}^\ell  (\ell+1)
   \left(\begin{matrix}
n+1-\ell\\
n+1-s
\end{matrix}\right)
\left(\begin{matrix}
\ell\\
k
\end{matrix}\right) 
D^{s-k} \tau_s(z)
  D^k\mathscr{S}_+(n-s+1,z)   \cr
 &=-\f{\kappa^2}{ 4}(-)^{s}  \sum_{k=0}^s 
(s-k+1 ) 
 \left(\begin{matrix}
n+3\\
k
\end{matrix}\right)
  D^k \tau_s(z)
  D^{s-k}\mathscr{S}_+(n-s+1,z),   \cr
  \end{split}
\ee
where we used 
\be
&\sum_{\ell=k}^s 
 (\ell+1)
   \left(\begin{matrix}
n+1-\ell\\
n+1-s
\end{matrix}\right)
\left(\begin{matrix}
\ell\\
k
\end{matrix}\right)\cr
&=
-\f{(1 + k)(n+3)  \Gamma(k - n) \Gamma(n-k+1 )}{ (n+k-s+3) (s-k)! \Gamma(n-s+2)}
\f{\Gamma(  s- n -k-2)}{\Gamma(s-n-1  )\Gamma(-n-2)}\cr
&=
(-)^{s+k+1} \f{(1 + k)(n+3)  }{ (n+k-s+3) (s-k)! }
\f{\Gamma(  s- n -k-2)}{\Gamma(-n-2)}\cr
&=-(1 + k) 
 \left(\begin{matrix}
n+3\\
s-k
\end{matrix}\right)\,.
\ee


\subsection{Higher order charges}\la{App:Qk}

Starting again from \eqref{qren}, the renormalized charge aspects at a general order $k$ are  given by
\be
\hat q^k_s(u,z)&=\frac12 \p_u^{-1} \sum_{\ell=k}^{s} \sum_{m=0}^{s-\ell} (\ell+1)  \frac{(-u)^{s-\ell}}{(s-\ell)!} 
   \left(\begin{matrix}
s-\ell\\
m
\end{matrix}\right)
D^{m} C (u,z)  D^{s-\ell-m} \cQ^{k-1}_{\ell-2}(u,z) \cr
&=\frac12 \p_u^{-1}\sum_{m=0}^{s-k} \f{(-u)^m}{m!} D^{m} C (u,z)
\sum_{\ell=k-2}^{s-m-2}  (\ell+3)  \frac{(-u)^{s-\ell-m-2}}{(s-\ell-m-2)!} 
  D^{s-\ell-m-2} \cQ^{k-1}_{\ell}(u,z) \cr
&=\frac12 \p_u^{-1}\sum_{m=0}^{s-k} \f{(-u)^m}{m!} D^{m} C (u,z)
\left[3 \hat q^{k-1}_{s-m-2}(u,z)+
\sum_{\ell=k-2}^{s-m-2}  \ell   \frac{(-u)^{s-\ell-m-2}}{(s-\ell-m-2)!} 
  D^{s-\ell-m-2} \cQ^{k-1}_{\ell}(u,z)
  \right]\,,\cr
\ee
where in the second line we used the relation \eqref{Qk}. 
We  now use the simple identity
\be
\f{\ell}{(s-m-2-\ell )!}=-\f{1}{(s-m-\ell -3)!} +\f{(s-m-2)}{(s-m-2-\ell )!}
\ee
to rewrite
\be
&\sum_{\ell=k-2}^{s-m-2} 
   \frac{\ell (-u)^{s-m-2-\ell}}{(s-m-2-\ell)!}
 D^{s-m-2-\ell} \cQ^{k-1}_{\ell} (u,z)\cr
&=
-\sum_{\ell=k-2}^{s-m-3} 
   \frac{ (-u)^{s-m-2-\ell}}{(s-m-\ell -3)!}
 D^{s-m-2-\ell} \cQ^{k-1}_{\ell} (u,z)
 +(s-m-2)\hat q^{k-1}_{s-m-2}(u,z)\cr
 &=uD\hat q^{k-1}_{s-m-3}(u,z) +(s-m-2)\hat q^{k-1}_{s-m-2}(u,z)\,,
\ee
and finally arrive at
\be
\hat q^k_s(u,z)&=\frac12 \p_u^{-1}\sum_{m=0}^{s-k} \f{(-u)^m}{m!} D^{m} C (u,z)
\left[(s-m+1) \hat q^{k-1}_{s-m-2}(u,z)+
uD\hat q^{k-1}_{s-m-3}(u,z)
\right]\,.
\ee

\bibliographystyle{bib-style2.bst}
\bibliography{biblio-w.bib}

\end{document}